\documentclass[nofootinbib,preprintnumbers,superscriptaddress,notitlepage]{revtex4}

\usepackage{epsfig}
\newcommand{\beq}{\begin{equation}}
\newcommand{\eeq}{\end{equation}}
\def\bea{\begin{eqnarray}}
\def\eea{\end{eqnarray}}

\begin{document}
\title{The evolution of neutrino masses and mixings in the 5D MSSM}

\author{A.~S.~Cornell}
\email[Email: ]{alan.cornell@wits.ac.za}
\affiliation{National Institute for Theoretical Physics; School of Physics, University of the Witwatersrand,
Wits 2050, South Africa}
\author{Aldo Deandrea}
\email[Email: ]{deandrea@ipnl.in2p3.fr}
\affiliation{Universit\'e de Lyon, F-69622 Lyon, France; Universit\'e Lyon 1, CNRS/IN2P3, UMR5822 IPNL, F-69622 Villeurbanne Cedex, France}
\author{Lu-Xin~Liu}
\email[Email: ]{luxin.liu9@gmail.com}
\affiliation{National Institute for Theoretical Physics; School of Physics, University of the Witwatersrand,
Wits 2050, South Africa}
\author{Ahmad Tarhini}
\email[Email: ]{tarhini@ipnl.in2p3.fr}
\affiliation{Universit\'e de Lyon, F-69622 Lyon, France; Universit\'e Lyon 1, CNRS/IN2P3, UMR5822 IPNL, F-69622 Villeurbanne Cedex, France}

\begin{abstract}
We consider a five-dimensional Minimal Supersymmetric Standard Model compactified on a $S^1/Z_2$ orbifold, and study the evolution 
of neutrino masses, mixing angles and phases for different values of $\tan \beta$ and different radii of compactification. We consider the 
usual four dimensional Minimal Supersymmetric Standard Model limit plus two extra-dimensional scenarios: where all matter superfields 
can propagate in the bulk, and where they are constrained to the brane. We discuss in both cases the evolution of the mass spectrum,
the implications for the mixing angles and the phases. We find that a large variation for the Dirac phase is possible, which makes 
models predicting maximal leptonic CP violation especially appealing.
\end{abstract}


\date{June 26th, 2012}
\preprint{LYCEN 2012-01, WITS-CTP-100}
\maketitle


\section{Introduction}\label{sec:1}

\par In the last decade, our perspectives in the search for physics Beyond the Standard Model (BSM) have seen the development of theories with extra (compact) dimensions. TeV-scale extra-dimensional models have allowed us to build effective models which can be tested at the present and next generation of colliders, where these wide varieties of extra-dimensional models have been proposed to solve, or at least understand from a geometrical perspective, different theoretical problems arising in the Standard Model (SM) and in its four-dimensional space-time extensions.

\par Another ingredient which can be introduced, supersymmetry, and in particular its minimal low energy construction, the Minimal Supersymmetric Standard  Model (MSSM), has long been held to play a central role in BSM physics; even if at present there is no evidence at colliders for the supersymmetric partners of the SM particles. From a theoretical point of view, supersymmetry plays a key role in resolving many problems in the SM and BSM physics; from gauge coupling unification, to the hierarchy problem, to the building of realistic grand unification models. The combination of supersymmetry and the physics of extra-dimensions has the added bonus of stabilising the extra-dimensional theory from quantum fluctuations, as well as the extra-dimensions potentially providing a mechanism for supersymmetry breaking.  Indeed, four-dimensional supersymmetric models typically lack a simple mechanism for supersymmetry breaking, which the extra-dimensions may offer. We will discuss in the following a five-dimensional $\mathcal{N}=1$ supersymmetric model compactified on the $S^1/Z_2$ orbifold (5D MSSM) as a simple testing ground for the effects of the extra-dimension on the neutrino sector. A similar study, concerning the quark Yukawa couplings and the CKM matrix observables was performed in Ref.~\cite{Cornell:2011fw}, which contains also a more detailed presentation of the general features of the 5D MSSM model.

\par Neutrinos are generally taken to be massless in the SM, and the experimental evidence for nonzero neutrino masses implicit in the neutrino oscillations measurements, gives an important indication for physics BSM. In fact, neutrino masses are many orders of magnitude smaller than those of quarks and charged leptons. However, contrary to the small mixings in the quark sector, two of the lepton mixing angles are identified as being rather large, close to maximal. For an overview of the present knowledge of neutrino masses and mixings see Ref. \cite{Mohapatra:2005wg,Raidal:2008jk} and references therein. The most recent experimental evidence of this fact is the measurement of the $\theta_{13}$ mixing parameter by the Daya Bay and RENO experiments \cite{An:2012eh,Ahn:2012nd}. The implication of three sizable mixing angles are huge, and will surely boost the number of investigations of the neutrino mixings and phases in the near future; in particular, this result suggests the possibility of measuring leptonic CP violation. The neutrino sector seems, therefore, to continue to play a special role in understanding BSM physics.

\par Recall also that quark and lepton masses and mixing angles are free parameters in minimal extensions to the SM. The running of both quark and neutrino masses and mixings has been investigated extensively in the SM and various BSM extensions \cite{Cornell:2011fw, Babu:1987im, Liu:2009vh, Cornell:2010sz, Liu:2011gr, Chankowski:1993tx, Antusch:2001ck, Chankowski:1999xc, Casas:2003kh, Blennow:2011mp, Antusch:2003kp}. Also, as neutrino mixing angles show a pattern that is completely different from that of quark mixings, the relative wealth of the latest experimental data has motivated efforts on the theoretical side to understand the possible patterns of neutrino masses and mixings, and therefore to expose the underlying fundamental symmetries behind them. As such, understanding the evolution of these neutrino sector parameters will be critical as higher energy experiments probe this sector. With this in mind, starting from the 5D MSSM, this paper is organized as follows: In section \ref{sec:2} we introduce the dimension five operator (in four space-time dimensions) for the neutrino masses in the 5D MSSM. Then in section \ref{sec:3} we further develop the renormalization group equations for the neutrino mass matrix. In section \ref{sec:4} we shall explore and discuss the evolution properties and behaviours for neutrino masses and mixing angles, and section \ref{sec:5} is devoted to our conclusions.


\section{5D MSSM and neutrino masses}\label{sec:2}

\par In order to study qualitative features in a model independent way, an attractive and simple possibility is to use a low-energy effective theory formulation. This means that additional particles and symmetries present at higher energies are organized within a systematic expansion. We shall consider the heavy states arising from BSM physics to decouple at low energies. In that case, the degree of suppression of an operator in the low energy effective Lagrangian is characterized by its mass dimension. This general framework is valid also when discussing extra-dimensional theories. In the 5D MSSM, the Higgs superfields and gauge superfields always propagate into the fifth dimension. However, different possibilities of localisation for the matter superfields can be studied. We shall consider the two limiting cases of superfields with SM matter fields all in the bulk or all superfields containing SM matter fields restricted to the brane. When all fields propagate in the bulk, the action for the matter fields $\Phi_i$ is \cite{Deandrea:2006mh}:
\begin{eqnarray}
S_{matter} &=& \int\mbox{d}^8z\mbox{d}y\left\{ \bar{\Phi}_i\Phi_i + \Phi^c_i\bar{\Phi}^c_i + \Phi^c_i\partial_5\Phi_i \delta(\bar{\theta}) - \bar{\Phi}_i\partial_5\overline{\Phi}_i^c\delta(\theta) \right. \nonumber \\
& & \left. \hspace{1.7cm} + \tilde{g}(2\bar{\Phi}_iV\Phi_i - 2\Phi_i^cV\bar{\Phi}^c_i + \Phi^c_i\chi \Phi_i\delta(\bar{\theta}) + \bar{\Phi}_i\bar{\chi}\bar{\Phi}^c_i\delta(\theta)) \right\} \; .
\label{eq:matterbulk}
\end{eqnarray}
Similarly, when all superfields containing SM fermions are restricted to the brane, the part of the action involving only gauge and Higgs fields is not modified, whereas the action for the superfields containing the SM fermions becomes:
\begin{equation}
S_{matter} = \int\mbox{d}^8z\mbox{d}y \delta(y) \left\{ \bar{\Phi}_i\Phi_i + 2\tilde{g}\bar{\Phi}_iV\Phi_i \right\} \; .\label{matterbrane}
\end{equation}
Neutrinos, being SM matter fields, will therefore be considered either localised or propagating in the bulk (in these two different scenarios). However, as the neutrino masses are roughly six orders of magnitude smaller than the other light SM fermions, and if neutrinos are Majorana particles, such a small mass could be understood if there is new physics beyond the electroweak scale. The lowest order operator, which generates Majorana neutrino masses after electroweak symmetry breaking (EWSB), is the lepton-number violating Weinberg operator \cite{Weinberg:1979sa}. This lowest order operator (appearing with dimension $d=5$ in four space-time dimensions) can be written as:
\begin{equation}
-\frac{\tilde{k}_{ij}}{4M}(\bar{L}^{ci}_{\alpha} \epsilon^{\alpha \beta} \phi_{\beta})(L^{j}_{\delta} \epsilon^{\delta \gamma} \phi_{\gamma})+h.c. \; ,
\label{eq:5dop}
\end{equation}
where $L$ and $\phi$ are the lepton and the Higgs doublet fields. $M$ is the typical heavy energy scale for the range of validity of the low-energy effective theory. An operator of this type can be generated, for instance, by the usual see-saw mechanism. In which case the heavy scale $M$ can be identified with the mass of the heavy right-handed neutrino. After EWSB the Higgs acquires a vacuum expectation value (vev) and the operator in Eq.~(\ref{eq:5dop}) gives a Majorana mass term for the neutrinos. In the context of the MSSM it can be written in the form:
\begin{equation}
-\frac{\tilde{k}_{ij}}{4M}(L^{i}_{\alpha} \epsilon^{\alpha \beta} H^{u}_{\beta})(L^{j}_{\delta} \epsilon^{\delta \gamma} H^{u}_{\gamma})\; ,
\label{eq:5dmssm}
\end{equation}
where $L$ and $H^{u}$ are the lepton and up-type Higgs doublet chiral superfields respectively. This operator is crucial for the study of neutrino masses and mixings, where renormalisation group equations for this effective operator have been derived in the context of the four-dimensional SM and MSSM \cite{Chankowski:1993tx,Antusch:2001ck}.

\par For $\mathcal{N}=1$, $D=4$ supersymmetry the superfield formalism is well established: superfields describe quantum fields and their superpartners as well as auxiliary fields as a single object, thus simplifying considerably the notations and the calculations. A similar formulation for a 5D vector superfield has been developed in Ref.~\cite{Mirabelli:1997aj}, and the superfield formulation for vector and matter supermultiplets can be found in Refs.~\cite{ArkaniHamed:2001tb,Hebecker:2001ke}. A detailed description of these techniques in the covariant formalism, together with the corresponding Feynman rules, is described in Ref.~\cite{Flacke:2003ac}.

\par In Eq.~(\ref{eq:5dmssm}) the neutrinos acquire Majorana masses through the dimension five operator. An extension to the compactified 5D MSSM was considered in Ref.~\cite{Deandrea:2006mh}, and we shall use a similar formalism here. That is, for the energy range below the mass scale of the heavy right-handed neutrino, we assume that the effective theory is just the effective 5D MSSM with $S^1/Z_2$ compactification, and the energy dependence of the effective neutrino mass matrix below that scale is then described by its renormalization group equation. Note that in our model the Yukawa couplings in the bulk are forbidden by the 5D $\mathcal{N}=1$ supersymmetry. However, they can be introduced on the branes, which are 4D subspaces with reduced supersymmetry. We will write the following interaction terms, called brane interactions, containing Yukawa-type couplings:
\begin{equation}
S_{brane}=\int\mbox{d}^8z\mbox{d}y\delta(y)\left\{ \left(\frac{1}{6} \tilde{\lambda}_{ijk}\Phi_i\Phi_j\Phi_k - \frac{\tilde{k}_{ij}}{4 M} L_iH_uL_jH_u\right)\delta(\bar{\theta}) + h.c. \right\} \; . \label{eq:5brane}
\end{equation}
The last term corresponds to the effective neutrino mass operator, with dimensional coupling $\tilde{k}_{ij}$ whose evolution and (that of the Majorana mass term for neutrinos) we shall now calculate.


\section{$k$ coupling evolution}\label{sec:3}

\par In the case that the neutrino superfields propagate in the bulk, it is straight forward for us to write down the following terms related to the neutrino masses, where we include terms related to both Majorana and Dirac masses:
\beq
\frac{{{\tilde{k}_{ij}}}}{4M}L_i^0H_u^0L_j^0H_u^0 + \frac{{{\tilde{k}_{ij}}}}{4M} \cdot 2\sqrt 2 L_i^nH_u^0L_j^0H_u^0+ h. c. + \frac{n}{R}\left({L ^{c(n)}}{L ^n} + {{\bar L }^{(n)}}{{\bar L }^{cn}}\right) \; . \label{neutrinomass}
\eeq
In which case the first two terms are from the $F$ terms of the dimension five operator in Eq.~(\ref{eq:5brane}) along with its Kaluza-Klein expansion in the bulk space, and the last term is from the $F$ term related to the gauge covariant derivative in Eq.~(\ref{eq:matterbulk}). $R$ is the compactification radius of the 5D MSSM.

\par In the present case we consider the effective neutrino mass operator with dimensional coupling $\tilde{k}_{ij}$; after spontaneous symmetry breaking, the Majorana neutrino masses can be written as $m_\nu \equiv k v^2 sin^2 \beta$ ($v$ being the vev of the Higgs field and $\tan \beta$, the ratio of the vevs of our two Higgs doublets) and $k=\tilde{k}_{ij}/(2 M \pi R)$ for bulk propagating, and $k=\tilde{k}/(2M)$ for brane localised matter superfield scenarios respectively. For simplicity we have the mass matrix for one generation of neutrinos as follows:
\beq
\left( {\begin{array}{ccc}
{m_\nu}&2\sqrt{2}m_\nu &0\\
{2\sqrt{2}m_\nu}&0&{\displaystyle \frac{n}{R}}\\
0&{\displaystyle \frac{n}{R}}&0
\end{array}} \right)\; ,
\eeq
with the basis being $\nu^T=(\nu_L,\nu_L^n,\nu_R^{c(n)})$. By using the inverse iteration method, and up to first order corrections, we observe that the three active neutrino flavour eigenstates $\nu_{\alpha}$ are related to their mass eigenstates $\nu_i$ through the transformation $\nu_ {\alpha} \approx U_{PMNS}(1+O(R^2m_{\nu}^2))\nu_i$, where we have chosen to work in the basis where the charged lepton Yukawa coupling matrix is diagonal, and the neutrino mixing matrix $U_{PMNS}$, up to a diagonal phase matrix, stems from the diagonalisation of the Majorana neutrino mass matrix $m_\nu$. That is
\begin{equation}
U^{T}_{PMNS}m_\nu U_{PMNS} = diag(m_1, m_2, m_3) \; , \label{eq:PMNS}
\end{equation}
with $m_i$ being the minimal extension SM neutrino masses. For the large hierarchy between the compactification scale $1/R$ (which is usually taken at the TeV level for LHC physics) and the neutrino mass (eV level), we conclude $\nu_ {\alpha}= U_{PMNS}\nu_i$, up to a correction of order $10^{-24}$.

\par As in our previous works, the evolution of the Yukawa couplings were derived using standard techniques (see for example Appendix \ref{app:A} and Ref.~\cite{Deandrea:2006mh}). The evolution of $k$ can then be determined for the two scenarios we wish to consider along the same lines as we have done earlier \cite{Deandrea:2006mh}. As such, the beta function of $k$ at one loop for the MSSM is:
\begin{eqnarray}
(16 \pi^2)\beta_k &=& \alpha k + \left( [{Y_e}^T {Y_e}^*]k+k[{Y_e}^\dag Y_e] \right) C \nonumber\\
&=& \left( 6Tr(Y_u^\dag {Y_u}) - \frac{6}{5}g_1^2 - 6 g_2^2 \right) k + \left( [{Y_e}^T {Y_e}^*]k+k[{Y_e}^\dag Y_e] \right) 1 \; ,
\end{eqnarray}
where $C=1$ and $\alpha=6Tr(Y_u^\dag {Y_u}) - \frac{6}{5}g_1^2 - 6 g_2^2 = 6(y_t^2 + y_c^2 +y_u^2)- \frac{6}{5}g_1^2 - 6 g_2^2$.

\par We will use $S(\mu)=\mu R$, $t=\ln (\frac{\mu}{M_Z})$ and $S(t) = {e^t}{M_Z}R$, where at one loop for the case in which matter fields propagate in the bulk:
\begin{eqnarray}
(16 \pi^2)\beta_k &=& \alpha k + \left( [{Y_e}^T {Y_e}^*]k+k[{Y_e}^\dag Y_e] \right) C \nonumber\\
&=& \left( 6\pi \mu^2 R^2 Tr(Y_u^\dag {Y_u}) - (\frac{6}{5}g_1^2 +6 g_2^2)\mu R \right) k + \left( [{Y_e}^T {Y_e}^*]k+k[{Y_e}^\dag Y_e] \right) \pi \mu^2 R^2\; ,
\end{eqnarray}
where in this case $C = \pi S(t)^2$ and $\alpha = 6 \pi S(t)^2 Tr(Y_u^\dag Y_u) - (\frac{6}{5} g_1^2 + 6 g_2^2) S(t)$.

\par The beta function of $k$ at one loop for the brane localised matter fields case is then 
\begin{eqnarray}
(16 \pi^2)\beta_k &=& \alpha k + \left( [{Y_e}^T {Y_e}^*]k+k[{Y_e}^\dag Y_e] \right) C \nonumber\\
&=& \left( 6Tr(Y_u^\dag {Y_u}) - (\frac{9}{5}g_1^2 + 9 g_2^2)\mu R \right) k + \left( [{Y_e}^T {Y_e}^*]k+k[{Y_e}^\dag Y_e] \right) 2 \mu R \; ,
\end{eqnarray}
where now we have $C = 2 S(t)$ and $\alpha = 6 Tr(Y_u^\dag Y_u) - (\frac{9}{5} g_1^2 + 9 g_2^2) S(t)$.

\par From Eq.~(\ref{eq:PMNS}) and by inserting this into the beta function for $k$, one obtains the evolution for the neutrino mixing matrix due to the Kaluza-Klein modes. Furthermore, in order to derive the renormalization group running behaviour of the neutrino mixing parameters, we employ the standard parameterization, in which $U_{PMNS}$ is parameterized by three mixing angles and three CP-violating phases, see appendix \ref{app:0} for details. These equations are combined to arrive at the runnings for the leptonic mixing angles, as given in appendix \ref{app:B}, and where we have also included the phases in appendix \ref{app:C}.

Note that as a general remark, it is typical that when all fields are in the bulk that extra-dimensional effects are more important in a number of observables and, at the same time, the typical cut-off of the effective theory is lower with respect to the brane localised field model (as perturbative unitarity is violated at a lower scale and the strong running of many observables may drive the model outside the domain of perturbation theory for which the evolution equations can be applied). We will examine in detail the behaviour of the neutrino sector observables with this constraint in mind in the next section.


\section{Numerical results}\label{sec:4}

\par In this section we will apply the beta functions to the study of the evolution of the different couplings. Unless explicitly mentioned we refer to the {\it normal} hierarchy for the neutrino masses $m_3>m_2>m_1$. As there is some freedom in the choice of the mass values for the initial conditions, within the allowed range of $\Delta m_{sol}^2$ and $\Delta m_{atm}^2$, we have explicitly used in the case of a normal hierarchy the values for initial low energy conditions indicated at the end of Appendix \ref{app:0}.

\par The discussion of the fixed point structure for the mixing angles in the infrared is the same as the MSSM one \cite{Chankowski:1999xc,Casas:2003kh}, where the evolution equations for the neutrino masses and mixings are formally similar to the MSSM case since the beta functions have a similar structure. As expected $\tan\beta$ plays an important role as all the mixing angles and phases depend on $y_\tau$. However, the new degrees of freedom (the extra-dimensional fields giving rise to Kaluza-Klein excitations of the zero modes) become important at energies corresponding to their masses. In the following we study the evolution of the relevant parameters, such as $\Delta m_{sol}^2$, $\Delta m_{atm}^2$ and the angles and phases, as a function of the energy scale and of $\tan \beta$. Only some selected plots will be shown and we will comment on the other similar cases not explicitly shown.

\par In general, in the brane case, the evolution has the same form for the three masses $m_1$, $m_2$, $m_3$. This leads to a reduction of up to a factor of two for the masses at $t=6$ (for a large radius, $R^{-1}=1$ TeV) with respect to the MSSM values at low energies (smaller radii give a weaker effect as the Kaluza-Klein excitations contribute to the evolution equations at higher energies). This prediction is extremely stable and can be explained as the evolution of the masses is governed by the equation
\beq
\dot{m}_i = \frac{1}{16\pi^2} m_i \left( \alpha +  C_i y^2_\tau \right)\; ,
\eeq
where the coefficients $C_i$ induce a non-universal behaviour and the parameter $\alpha$ contains the up Yukawa couplings and the gauge coupling terms (detailed in Appendix \ref{app:B}). In contrast to the MSSM, the evolution in the brane case is completely dominated by the universal part. The essential point is that in the MSSM the positive contribution to $\alpha$, approximately $6y_t^2$, is of the same order as the negative contribution from the gauge part. In our case the gauge part has a large pre-factor $S(t)=e^t M_Z R$ with respect to the MSSM which makes it completely dominant compared to any other contribution. As energy increases, we can write:
\beq
\dot{m}_i \sim \frac{1}{16\pi^2} m_i\left[ 6y_t^2- \left(\frac{9}{5}g_1^2+9g_2^2\right) S(t)\right]< 0\; . \label{eq:branemassRGE}
\eeq
From this approximation we immediately see that all masses decrease with increasing energy and eventually tend to zero if the evolution equations can be trusted up to a high energy.

\par In the bulk case the evolution has again the same form for the three masses $m_1$, $m_2$, $m_3$, but the behaviour is the 
opposite as the masses increase at high energy because all matter fields propagate in the bulk and contribute to the evolution. In detail 
this can be seen by the fact that even if the gauge part gets a large pre-factor $S(t)$, the Yukawa part gets in this case a pre-factor 
$S(t)^2$, which changes the sign of the derivative with respect to the previous case:
\beq
\dot{m}_i \sim \frac{1}{16\pi^2} m_i\left[ \pi S(t)^2 6y_t^2- \left(\frac{6}{5}g_1^2+6 g_2^2\right) S(t)\right]>0\; . \label{eq:bulkmassRGE}
\eeq
From this approximation we see that all masses increase with increasing energy scale.

\par
\begin{figure}[htb]
\begin{center}
  \mbox{\epsfxsize=0.5\textwidth\epsffile{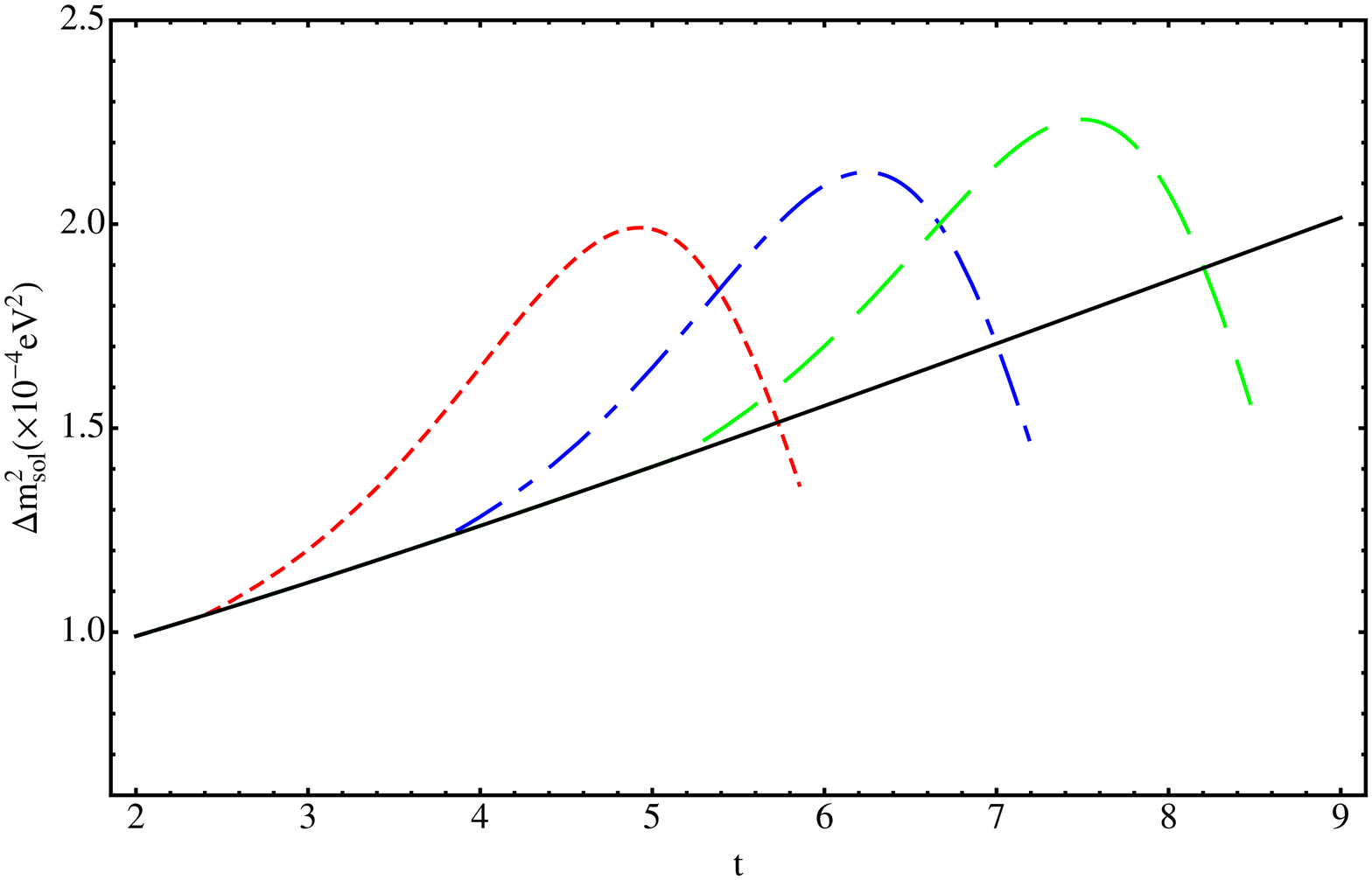} \epsfxsize=0.5\textwidth\epsffile{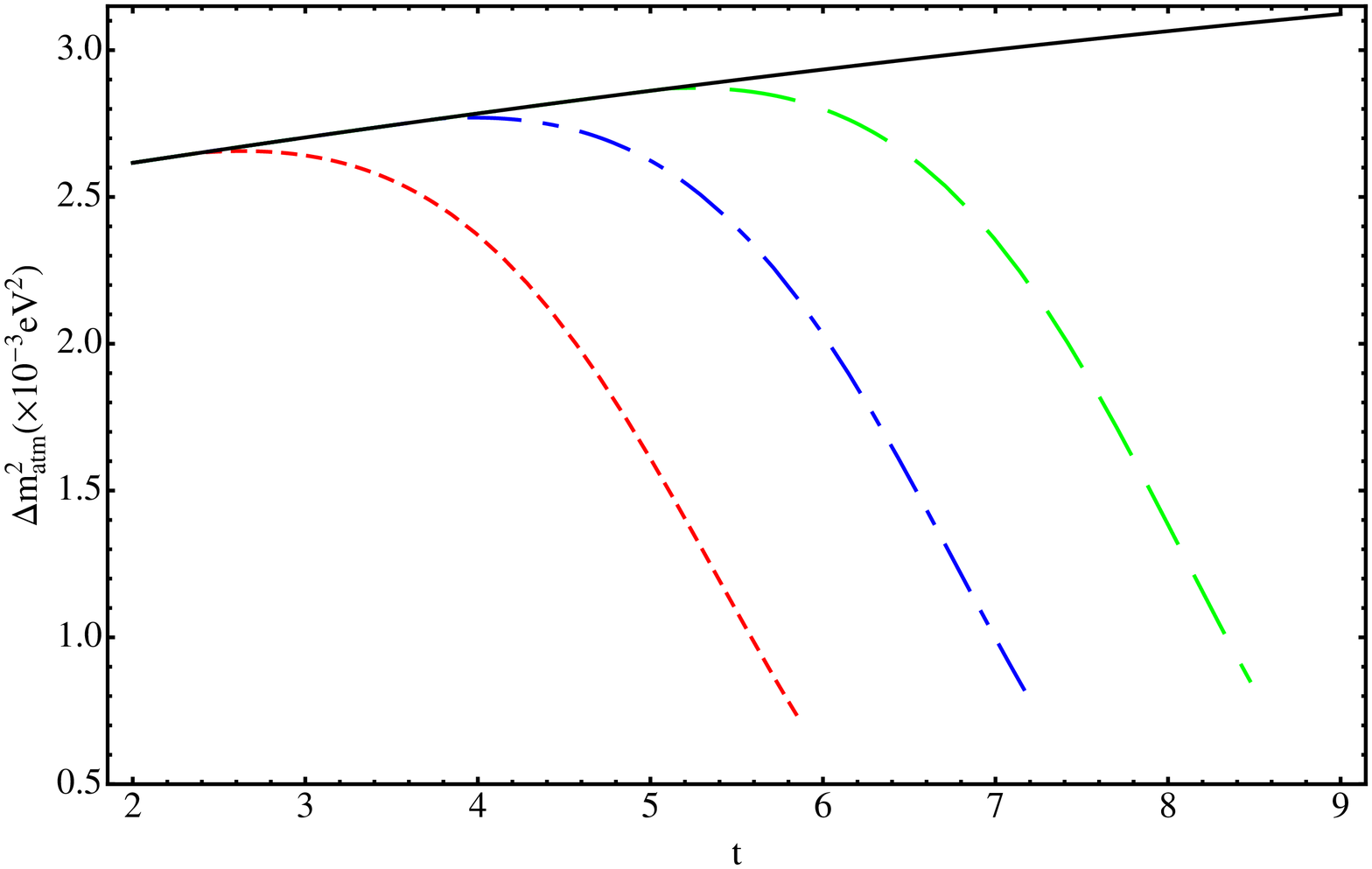}}
  \end{center}
\caption[]{\it Evolution of $\Delta m_{sol}^2$  (left panel) and  $\Delta m_{atm}^2$ (right panel)  as a function of the parameter $t=\ln (\mu/M_Z)$ with matter fields constrained to the brane. In this plot we have taken $\tan \beta = 30$. The black line is the MSSM evolution, the red (small dashes) is for $R^{-1}\sim 1$ TeV, the blue (dash-dotted) $R^{-1}\sim 4$ TeV, the green (large dashes) $R^{-1}\sim 15$ TeV.}
\label{fig:1}
\end{figure}

\begin{figure}[htb]
\begin{center}
  \mbox{\epsfxsize=0.5\textwidth\epsffile{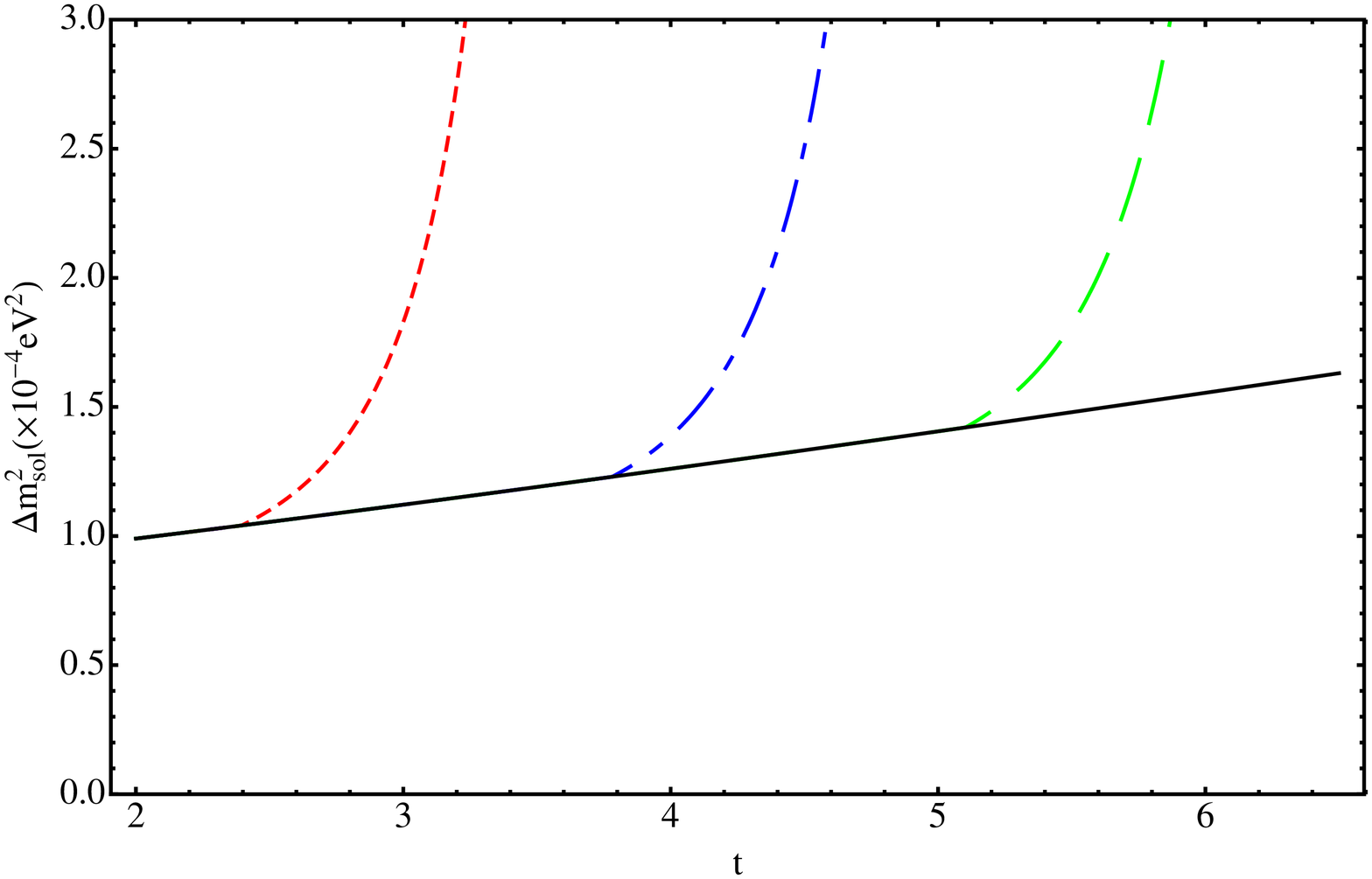} \epsfxsize=0.5\textwidth\epsffile{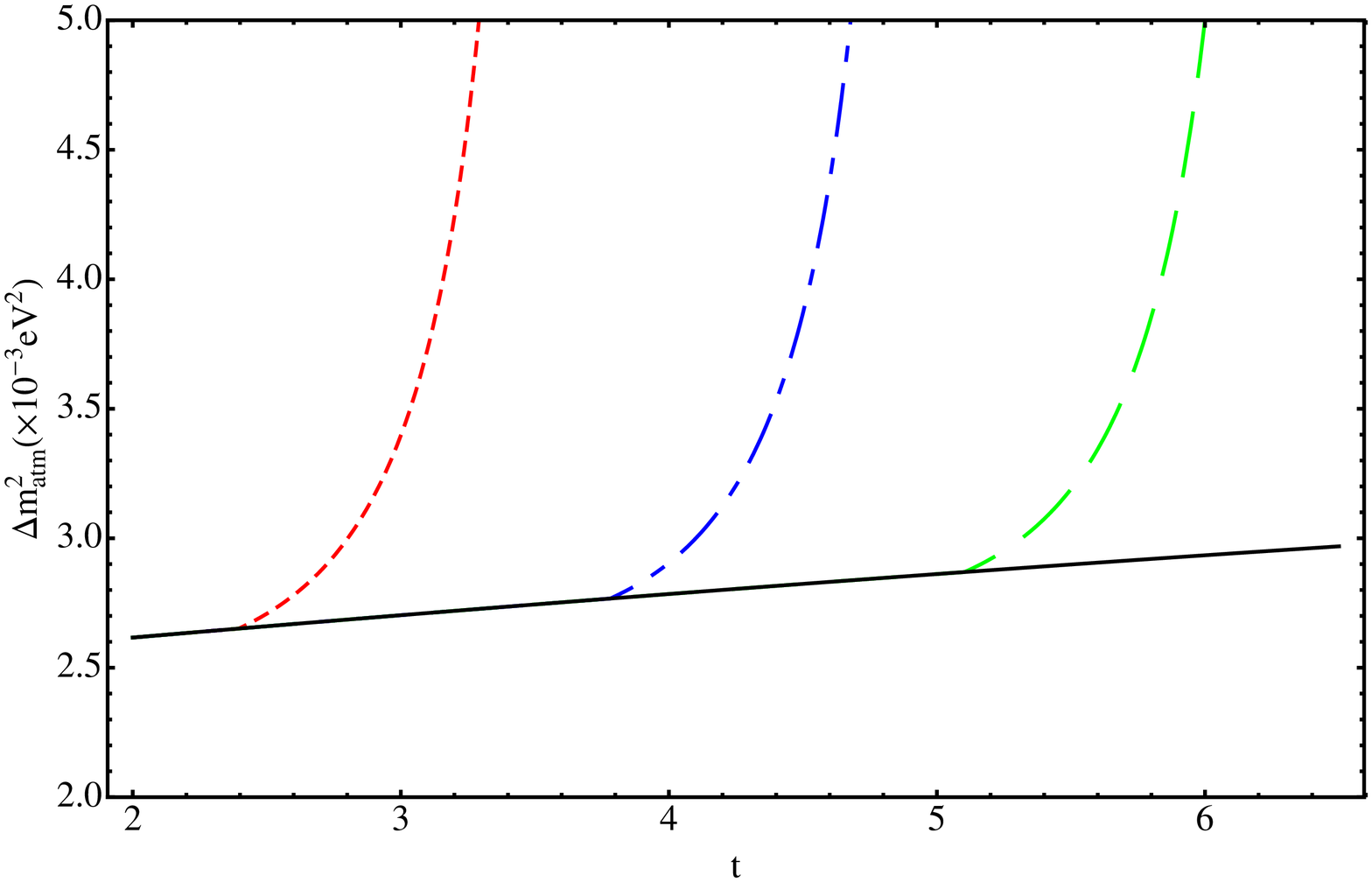}}
  \end{center}
\caption[]{\it Evolution of $\Delta m_{sol}^2$  (left panel) and  $\Delta m_{atm}^2$ (right panel) as a function of the parameter $t=\ln (\mu/M_Z)$ with matter fields in the bulk. In this plot we have taken $\tan \beta = 30$.
The black line is the MSSM evolution, the red (small dashes) is for $R^{-1}\sim 1$ TeV, the blue (dash-dotted) $R^{-1}\sim 4$ TeV, and the green (large dashes) $R^{-1}\sim 15$ TeV. The evolution is towards large values at high energies, the divergence signaling the breakdown of the effective theory.}
\label{fig:2}
\end{figure}

 The situation is more involved when analyzing the mass squared differences. We plot in Figs.~\ref{fig:1} and \ref{fig:2} the evolution of $\Delta m_{sol}^2$ and $\Delta m_{atm}^2$ both for the matter fields on the brane and for all fields in the bulk. In the brane case different behaviours as a function of the energy scale are possible as a relatively large interval in energy range is allowed for the effective theory. As explicitly illustrated in Fig.~\ref{fig:1}, the relevant radiative corrections controlled by the gauge fields in Eq.~(\ref{eq:branemassRGE}) become dominant as energy goes up, which tends to reduce mass splitting, and an approximately degenerate neutrino masses spectrum at the high energy scale $m_1 \approx m_2 \approx m_3$ becomes favourable. This is in contrast with the MSSM, where the neutrino mass splitting becomes large at an ultraviolet cut-off. Therefore, it is very appealing that the neutrino mass splitting at low energy could be attributed to radiative corrections resulting from a degenerate pattern at a high energy scale. In Fig.~\ref{fig:2}, the bulk case tends to a non-perturbative regime, where the unitarity bounds of the effective theory are reached much faster and only a much shorter running can be followed using the effective theory. As seen in Eq.~(\ref{eq:bulkmassRGE}), the quadratic terms related to $S(t)$ dominate during the fast evolution. As such, the neutrino mass splitting becomes even larger at a high energy scale.


\par
\begin{figure}[htb]
\begin{center}
  \mbox{\epsfxsize=0.5\textwidth\epsffile{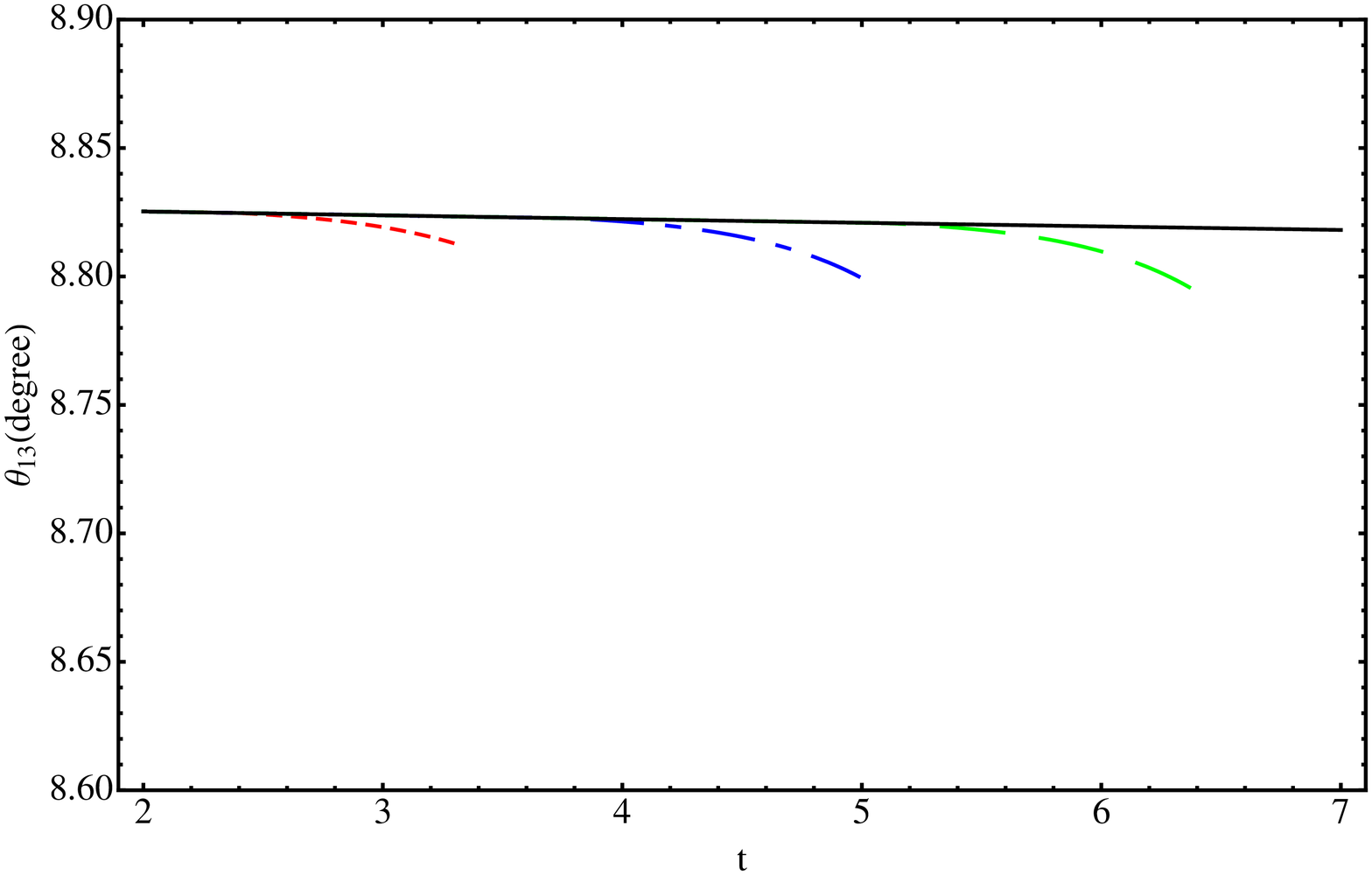} \epsfxsize=0.5\textwidth\epsffile{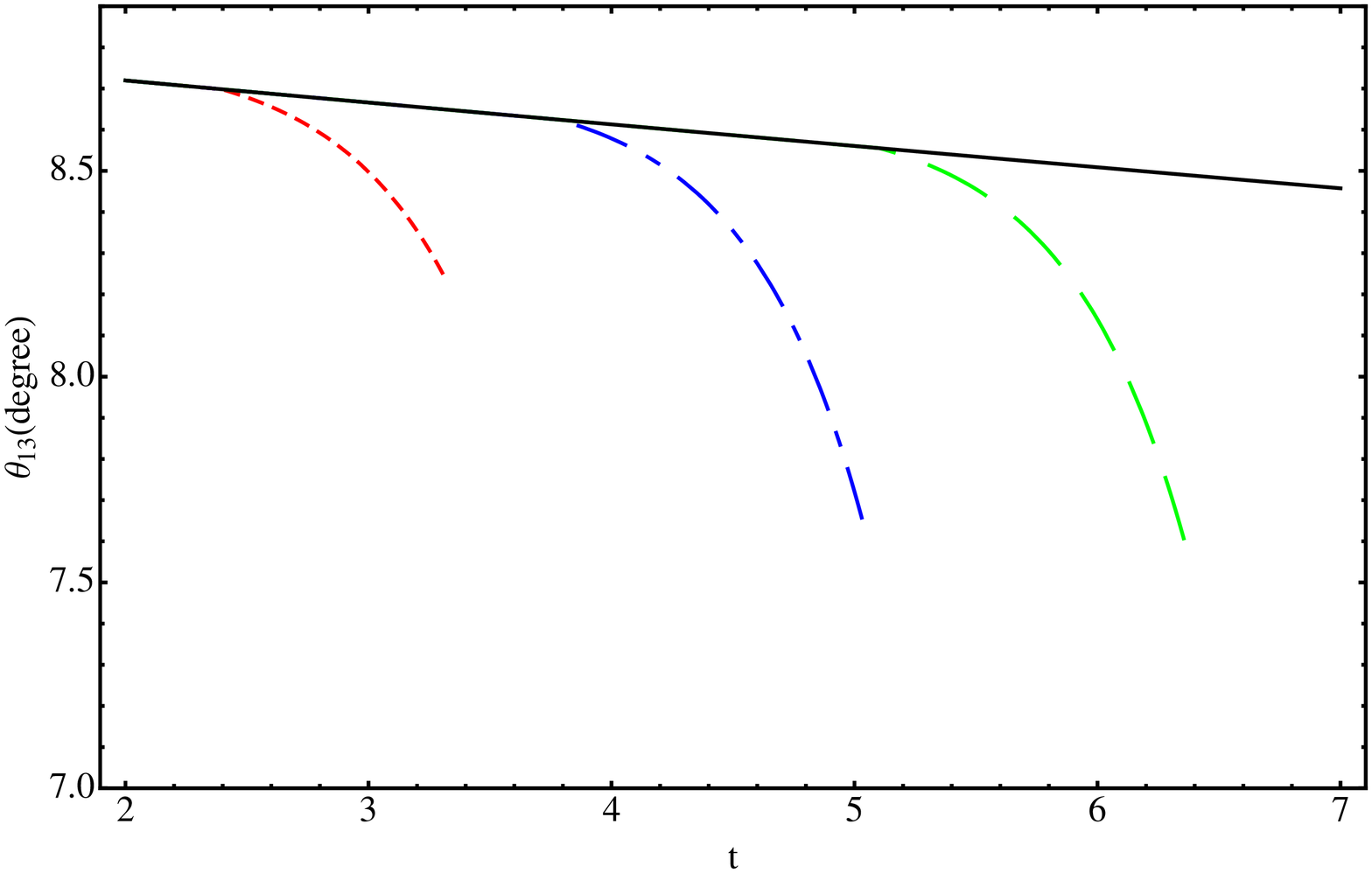}}
  \end{center}
\caption[]{\it Evolution of $\theta_{13}$  as a function of the parameter $t=\ln (\mu /M_Z)$ with matter fields in the bulk. In this plot we have taken $\tan \beta = 5$ ( left panel) and $\tan \beta = 30$ (right  panel) respectively. The black line is the MSSM evolution, the red (small dashes) is for $R^{-1}\sim 1$ TeV, the blue (dash-dotted) $R^{-1}\sim 4$ TeV, and the green (large dashes) $R^{-1}\sim 15$ TeV.}
\label{fig:3}
\end{figure}

\begin{figure}[htb]
\begin{center}
  \mbox{\epsfxsize=0.5\textwidth\epsffile{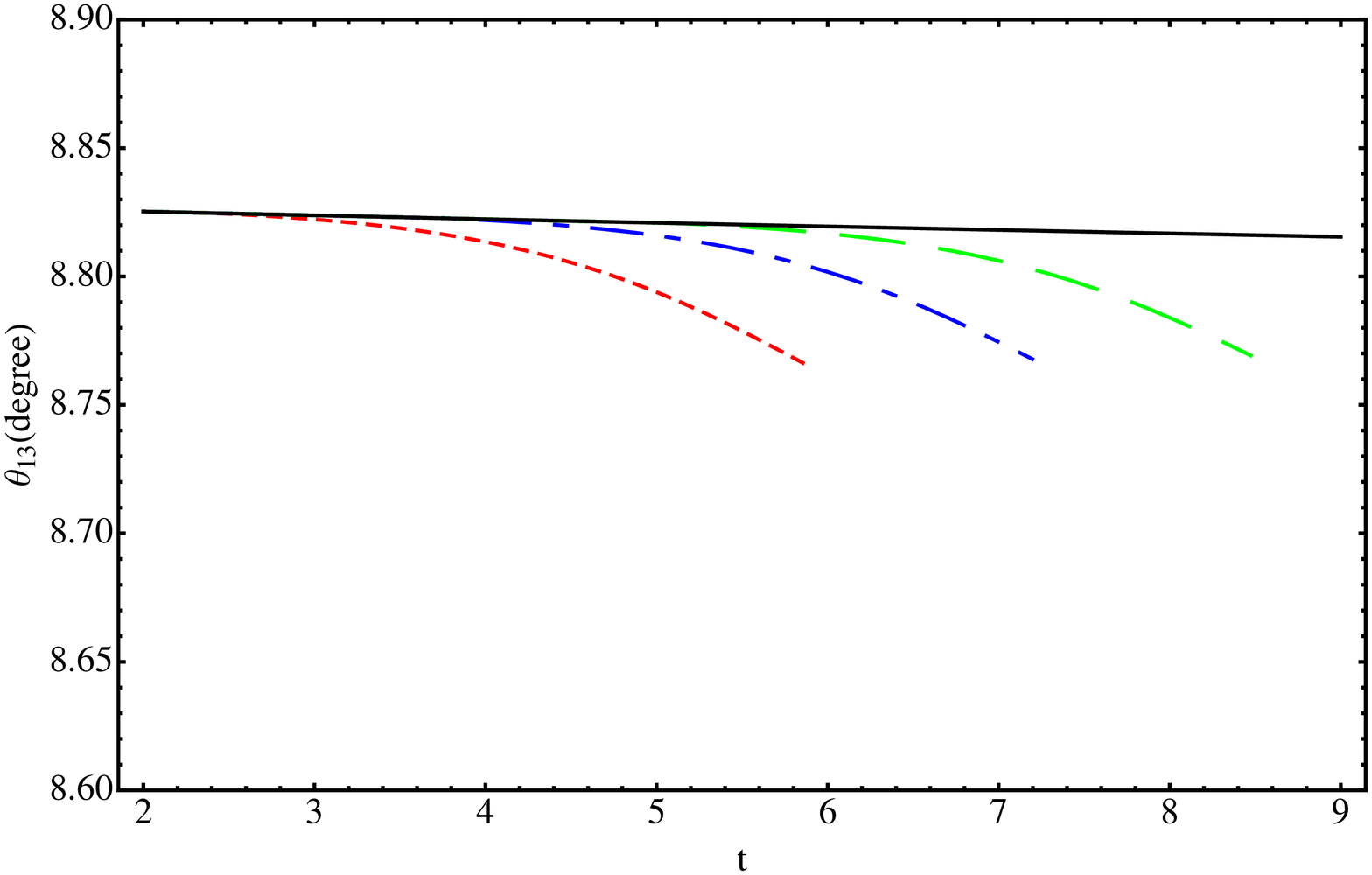} \epsfxsize=0.5\textwidth\epsffile{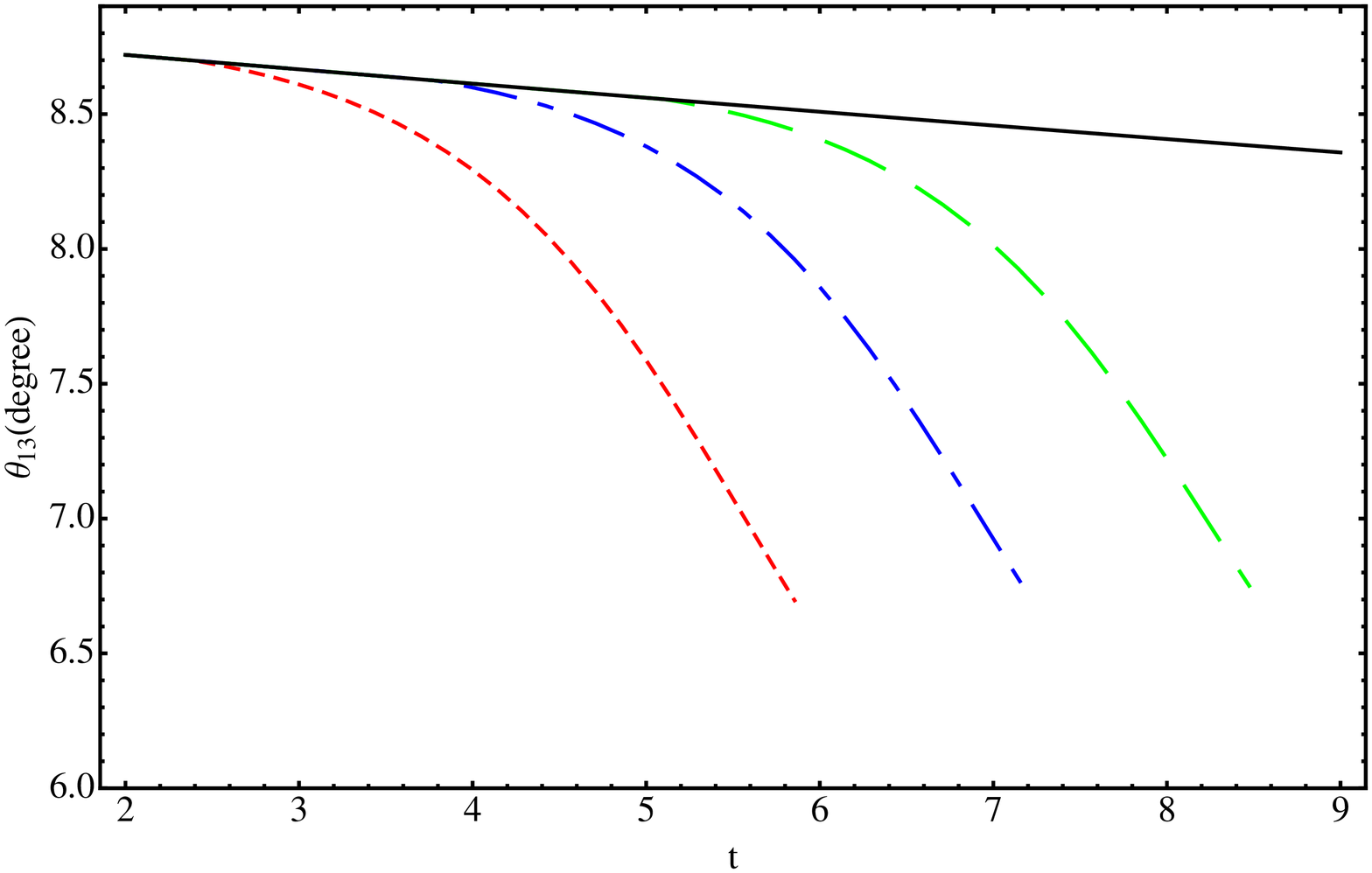}}
  \end{center}
\caption[]{\it Evolution of $\theta_{13}$ as a function of the parameter $t=\ln (\mu /M_Z)$ with matter fields constrained to the brane. In this plot we have taken $\tan \beta = 5$ ( left panel) and $\tan \beta = 30$ (right  panel) respectively. The black line is the MSSM evolution, the red one (small dashes) is for $R^{-1}\sim 1$ TeV, the blue (dash-dotted) $R^{-1}\sim 4$ TeV, and the green (large dashes) $R^{-1}\sim 15$ TeV.}
\label{fig:4}
\end{figure}

\begin{figure}[htb]
\begin{center}
  \mbox{\epsfxsize=0.5\textwidth\epsffile{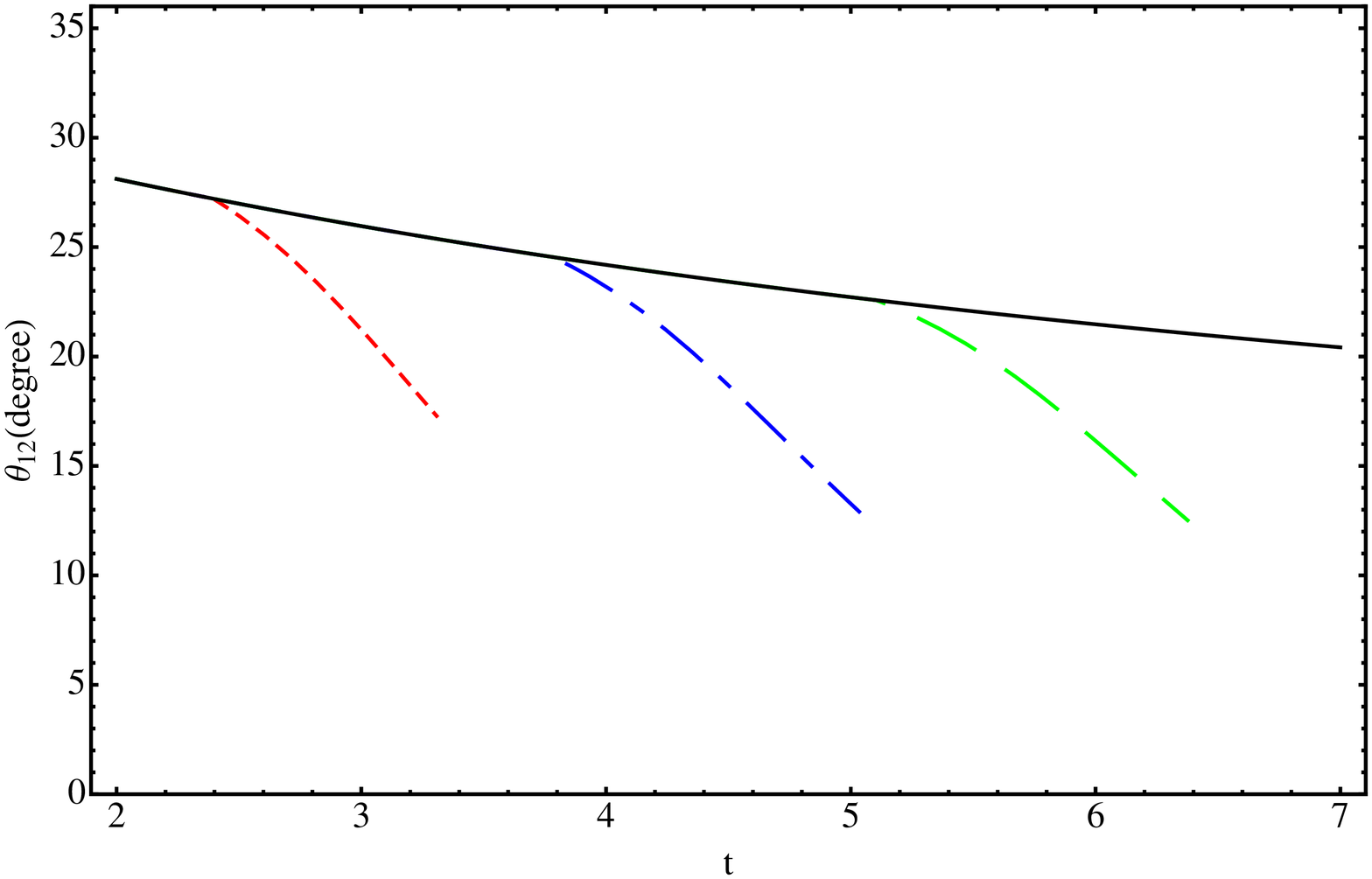} \epsfxsize=0.5\textwidth\epsffile{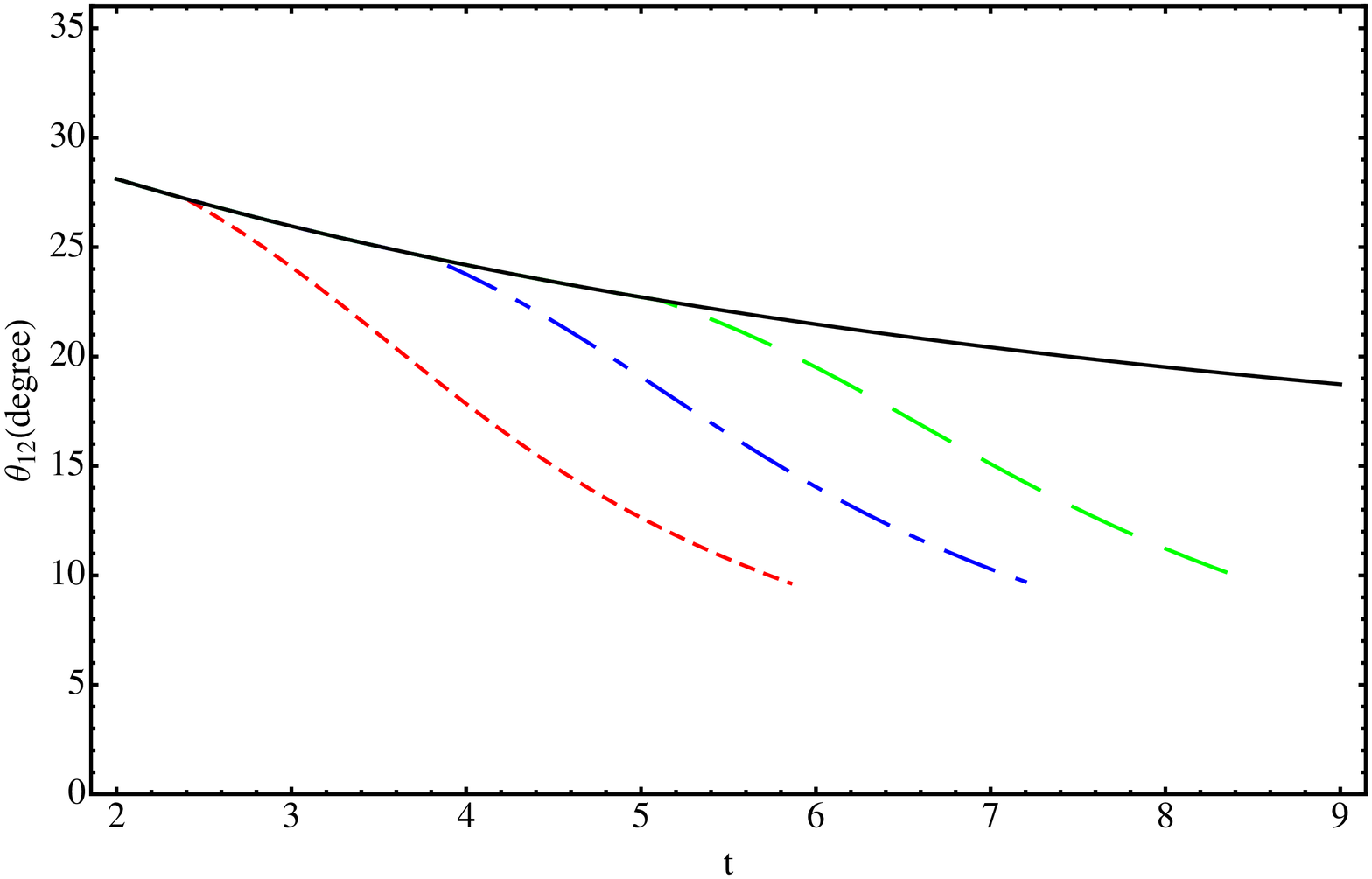}}
  \end{center}
\caption[]{\it Evolution of $\theta_{12}$ in the bulk (left panel) and on the brane (right panel) as a function of the parameter $t=\ln (\mu/M_Z)$. In this plot we have taken $\tan \beta = 30$. The black line is the MSSM evolution, the red one (small dashes) is for $R^{-1}\sim 1$ TeV, the blue (dash-dotted) $R^{-1}\sim 4$ TeV, and the green (large dashes) $R^{-1}\sim 15$ TeV.}
\label{fig:5}
\end{figure}

\begin{figure}[htb]
\begin{center}
  \mbox{\epsfxsize=0.5\textwidth\epsffile{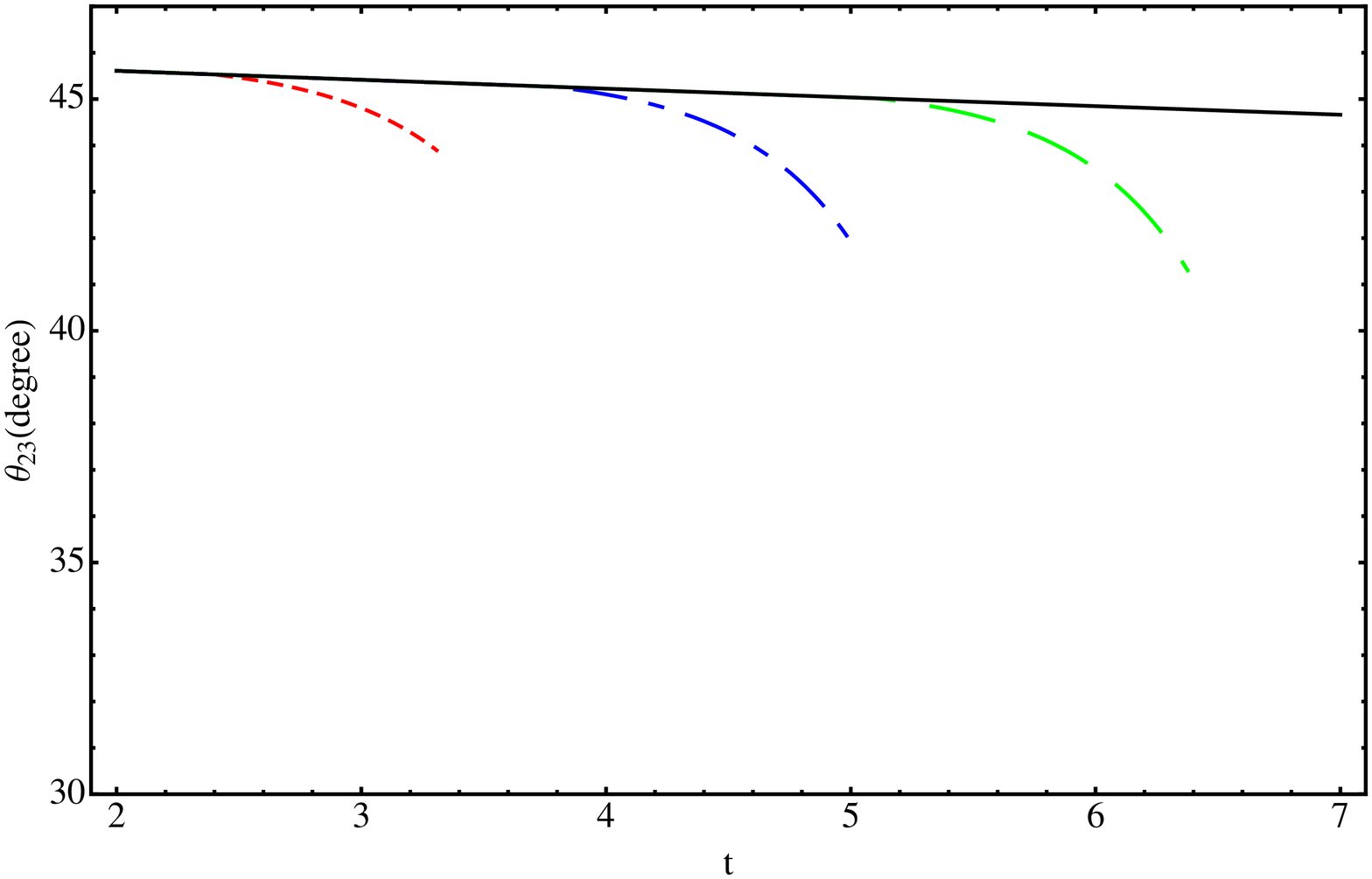} \epsfxsize=0.5\textwidth\epsffile{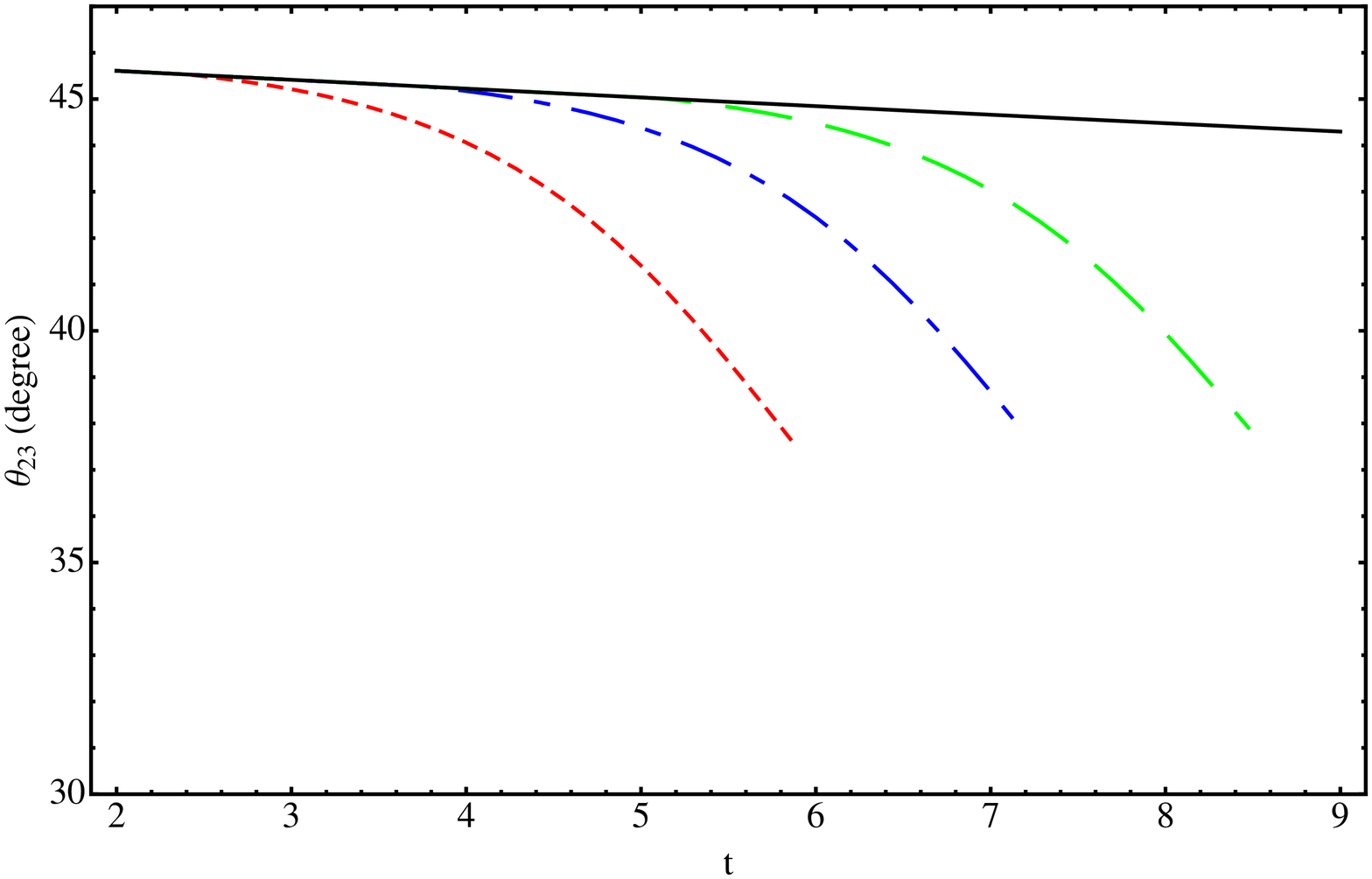}}
  \end{center}
\caption[]{\it Evolution of $\theta_{23}$ in the bulk (left panel) and on the brane (right panel) as a function of the parameter $t=\ln (\mu/M_Z)$. In this plot we have taken $\tan \beta = 30$. The black line is the MSSM evolution, the red one (small dashes) is for $R^{-1}\sim 1$ TeV, the blue (dash-dotted) $R^{-1}\sim 4$ TeV, and the green (large dashes) $R^{-1}\sim 15$ TeV.}
\label{fig:6}
\end{figure}

 Concerning the evolution of the mixing angles, as can be seen in Figs.~\ref{fig:3}--\ref{fig:6}, the largest effect is for $\theta_{12}$, with changes of more than 70\% possible for the brane localised matter field scenario. As observed, due to the large quadratic term of $S(t)$ in the beta function, the $\theta_{12}$ has a rapid and steep variation in the bulk case. However, for the brane case, it has a relatively longer evolution track with the $\theta_{12}$ then being pulled further down until the termination point (where the effective theory becomes invalid). In contrast, the running of $\theta_{13}$ and $\theta_{23}$ is much milder. As demonstrated in Figs. \ref{fig:3} and \ref{fig:4}, changes in the values of $\theta_{13}$ vary only a couple of degrees. For a larger value of $\tan \beta$ we have a relatively large Yukawa coupling to $\tau$, which leads to a large magnitude for its beta function, resulting in a relatively large variation during the evolution. However, a running to $\theta_{13} = 0$ cannot be observed in any situation. From the evolution behaviour of $\theta_{13}$, one can see that the renormalization group running effects or finite quantum corrections are almost impossible to generate $\theta_{13}=0$ at a high energy scale, even though the power law enhanced evolution is considered during the running. Therefore, for the tri-bimaximal mixing pattern \cite{Harrison:2002er}, in the current context with no other extreme conditions being taken into account, a slightly changed $\theta_{13}$ could not be accommodated during the whole range of the energy scale. Similar trajectories are also observed for $\theta_{23}$.


\par
\begin{figure}[htb]
\begin{center}
  \mbox{\epsfxsize=0.5\textwidth\epsffile{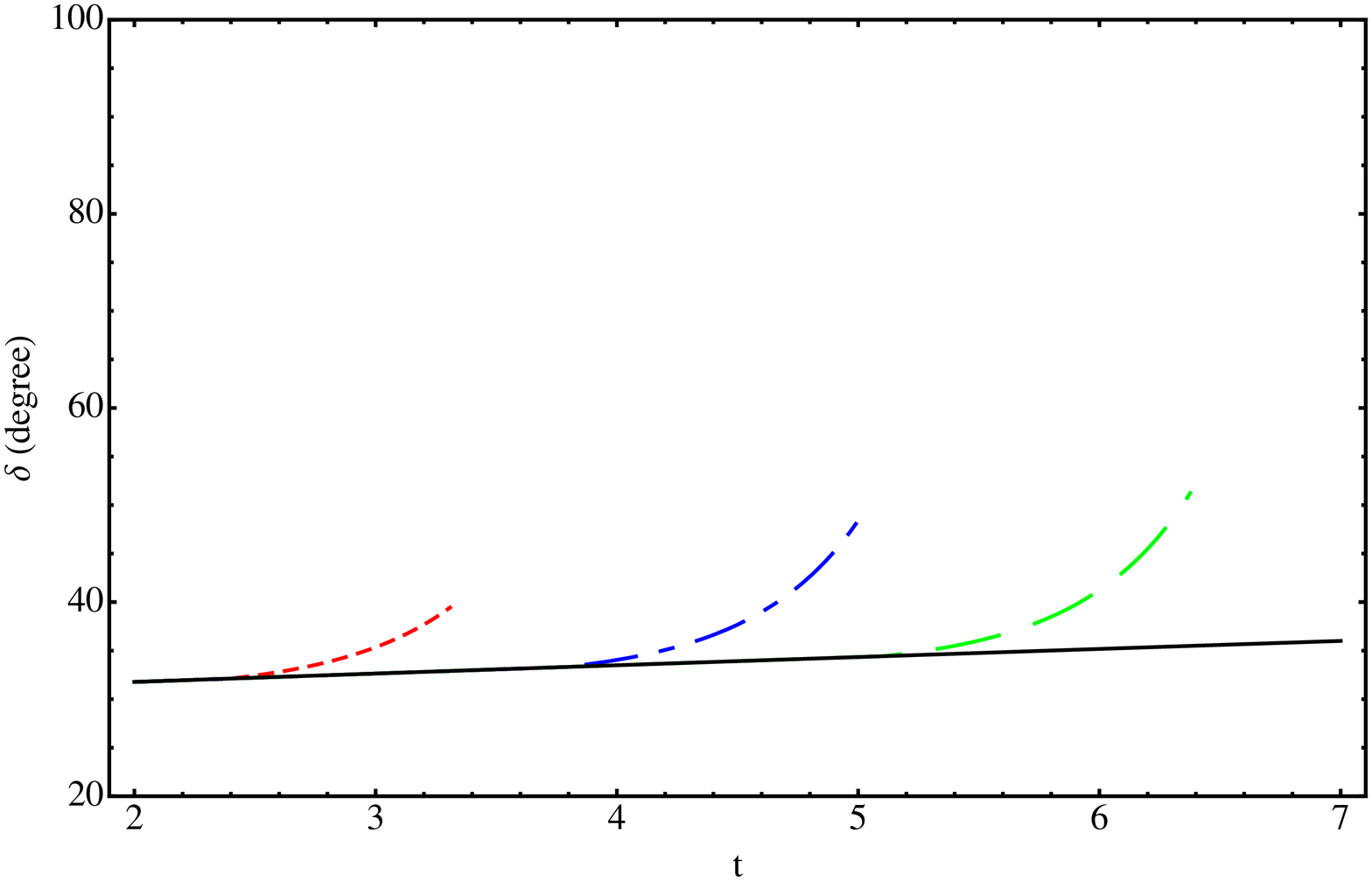} \epsfxsize=0.5\textwidth\epsffile{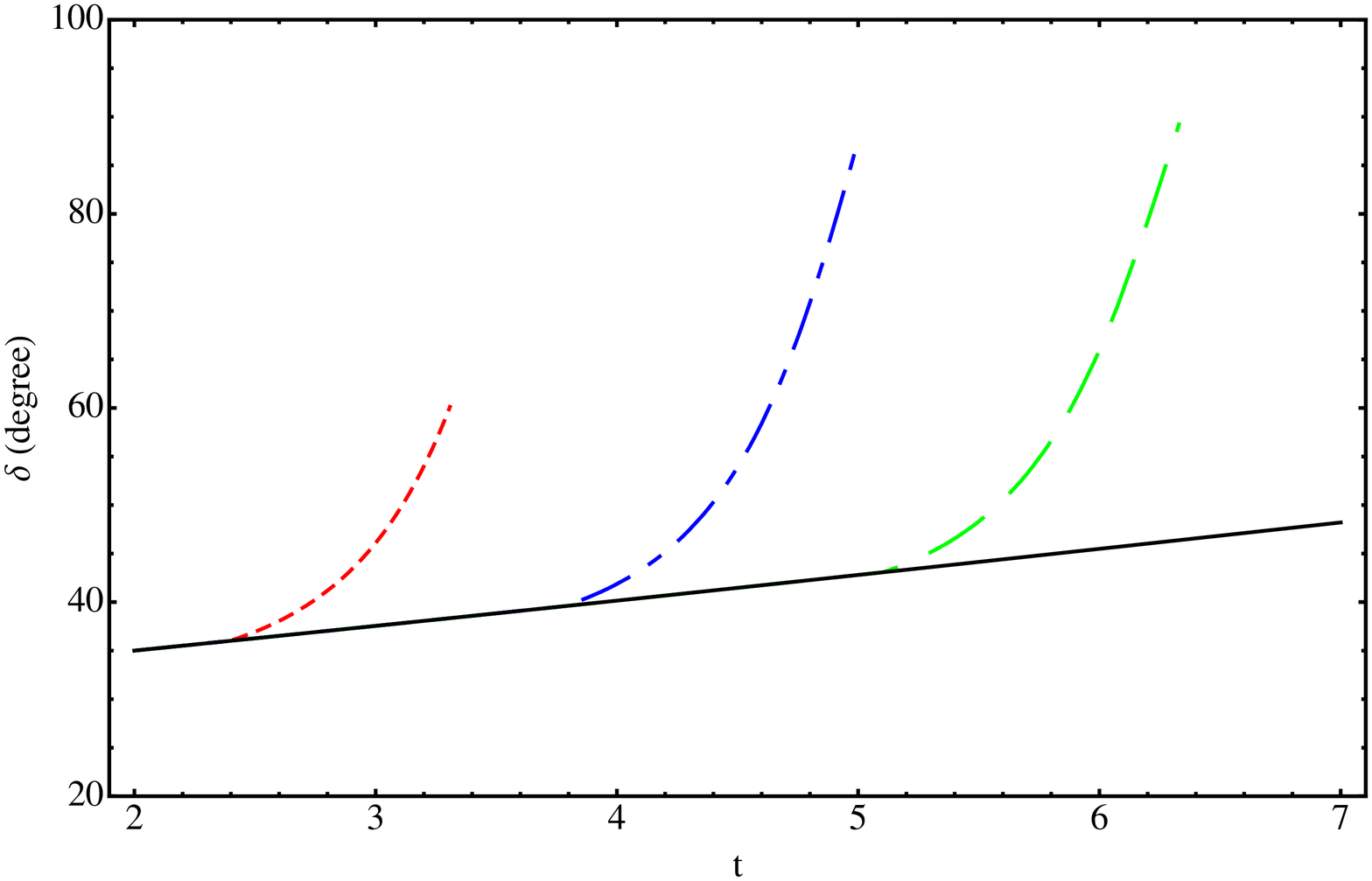}}
  \end{center}
\caption[]{\it Evolution of the phase $\delta$ as a function of the parameter $t=\ln (\mu/M_Z)$ with matter fields in the bulk. In this plot we have taken $\tan \beta = 30$ ( left panel) and $\tan \beta = 50$ (right  panel) respectively. The black line is the MSSM evolution, the red one (small dashes) is for $R^{-1}\sim 1$ TeV, the blue (dash-dotted) $R^{-1}\sim 4$ TeV, and the green (large dashes) $R^{-1}\sim 15$ TeV. The evolution is towards large values at high energies, the divergence signaling the breakdown of the effective theory.}
\label{fig:7}
\end{figure}

\begin{figure}[htb]
\begin{center}
  \mbox{\epsfxsize=0.5\textwidth\epsffile{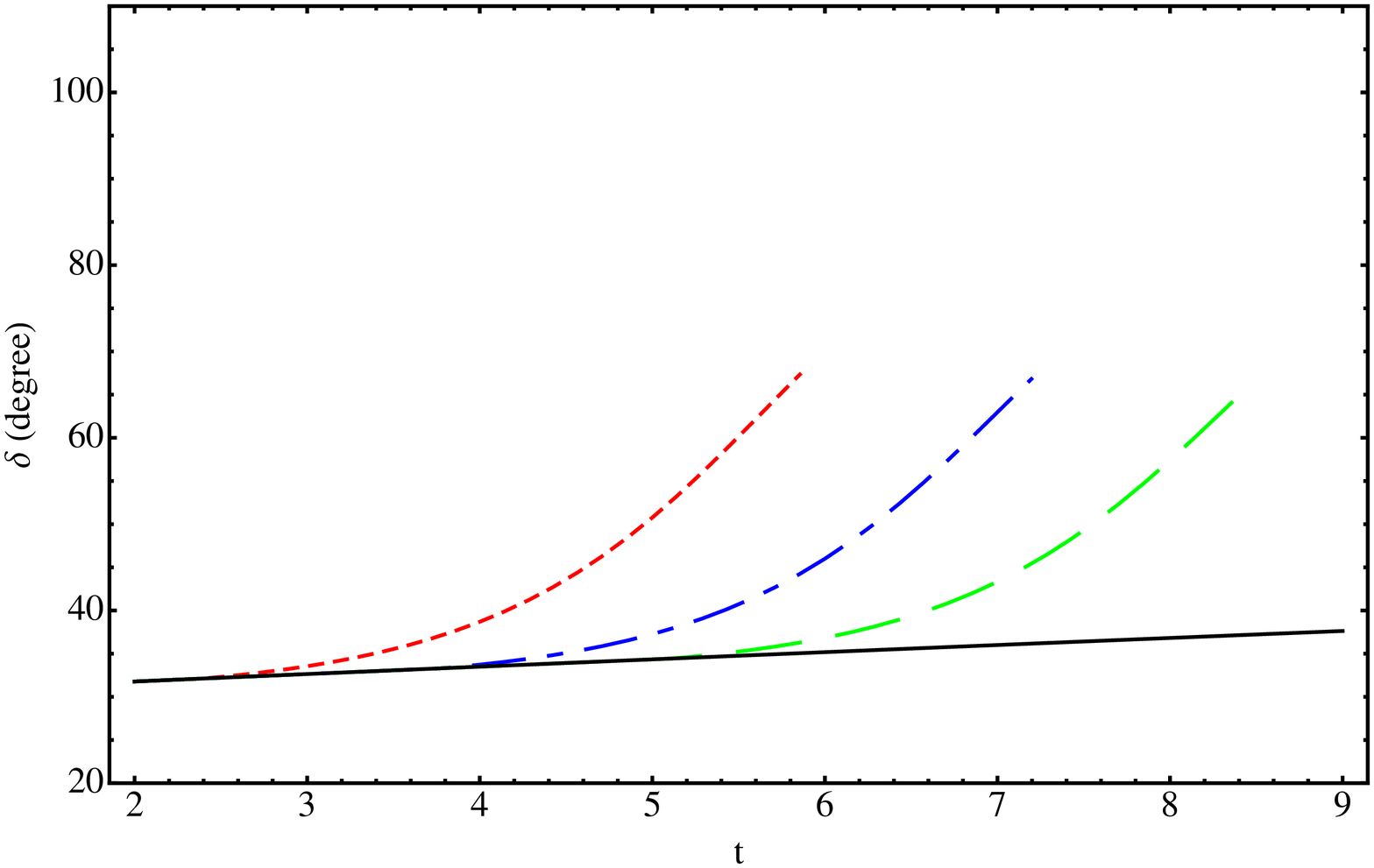} \epsfxsize=0.5\textwidth\epsffile{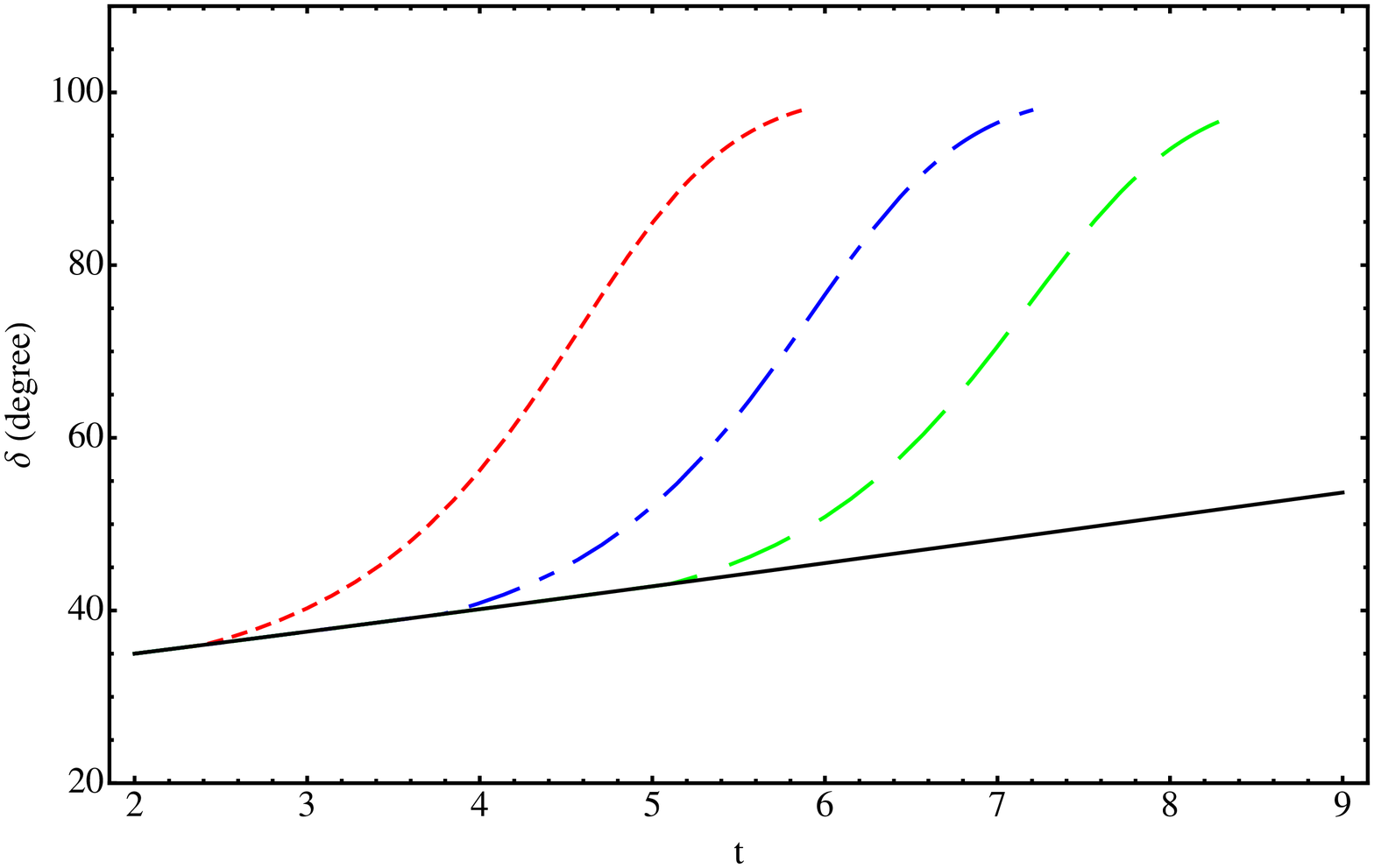}}
  \end{center}
\caption[]{\it Evolution of the phase $\delta$ as a function of the parameter $t=\ln (\mu/M_Z)$ with matter fields constrained to the brane. In this plot we have taken $\tan \beta = 30$ ( left panel) and $\tan \beta = 50$ (right  panel) respectively. The black line is the MSSM evolution, the red one (small dashes) is for $R^{-1}\sim 1$ TeV, the blue (dash-dotted) $R^{-1}\sim 4$ TeV, and the green (large dashes) $R^{-1}\sim 15$ TeV. The evolution is towards large values at high energies, the divergence signaling the breakdown of the effective theory.}
\label{fig:8}
\end{figure}

 Noting that the Dirac phase $\delta$ determines the strength of CP violation in neutrino oscillations. The runnings we include follow the general features presented in Figs.\ref{fig:7} and \ref{fig:8}, with large increases possible once the first Kaluza-Klein threshold is crossed. From these studies we have seen that the variation is bigger for high $\tan \beta$ with large changes appearing in the brane case when approaching the high energy scale. The recent results from the Daya bay and RENO reactor experiments have established a non zero values of $\theta_{13}$. Therefore, the leptonic CP violation characterized by the Jarlskog invariant $J \sim \sin {\theta _{12}}\cos {\theta _{12}}\sin {\theta _{23}}\cos {\theta _{23}}\sin {\theta _{13}}{\cos ^2}{\theta _{13}}\sin \delta$ becomes promising to be measured in the future long baseline neutrino oscillation experiments. As plotted, we can observe a relatively large evolution for the Dirac phase, even the maximum CP violation case $ \delta  = \frac{\pi }{2}$ could be achieved for relatively small input values. For leptogenesis related to the matter-antimatter asymmetry, we should note that the parameters entering the leptogenesis mechanism cannot be completely expressed in terms of low-energy neutrino mass parameters. Note that in some specific models the parameters of the PMNS matrix (which contains CP asymmetry effects) can be used \cite{Raidal:2008jk, Buchmuller:2003gz}. Here, the CP-violating effects induced by the  renormalization group corrections could lead to values of the CP asymmetries large enough for a successful leptogenesis, and the models predicting maximum leptonic CP violation, or where the CP-violating phase $\delta$ is not strongly suppressed, become especially appealing. Specific models with large extra dimensions in which leptogenesis is relevant at low scale can also be found in Ref. \cite{Gu:2010ye}.

\par As seen in Appendix \ref{app:B}, the running of the mixing angles are entangled with the CP-violating phases, and a more general discussion is beyond the scope of this paper. The phases $\phi_1$ and $\phi_2$ do not affect directly the running of the masses, while the phase $\delta$ has a direct effect on the size of ${dm}/{dt}$, although its importance is somewhat reduced by the magnitude of $\theta_{13}$. For further discussions of the correlation between these phases and mixing angles, refer to Refs.~\cite{Antusch:2003kp, Luo:2012ce} for details.


\par Finally, whilst the above results and analysis were for the {\it normal} hierarchy of neutrino masses, we did also consider an {\it inverted} hierarchy, where from an analysis of the equations presented in the Appendices we obtain the same features and results for neutrino mass runnings (though with different initial values at the $M_Z$ scale). As such, the figures for $\Delta m_{sol}^2$ and $\Delta m_{atm}^2$ remain unchanged. Possible changes in the angles and phases arise from the different signs for the $(m_j - m_i)/(m_j + m_i)$ terms present in each evolution equation, where the $\theta_{12}$ results remain approximately the same (given the relative ordering of masses in these two hierarchies), and the small runnings of $\theta_{13}$ and $\theta_{23}$ would be up rather than down (though as already discussed, these runnings are quite small).


\section{Conclusion}\label{sec:5}

\par In the present work we studied the neutrino sector in a minimal supersymmetric model with an extra-dimension using the updated value of $\theta_{13}$ and other mixing parameters. We have presented the evolution equations for the physical mixing angles, phases, $\Delta m_{sol}^2$ and $\Delta m_{atm}^2$, within two distinct scenarios, where a larger $\tan \beta$ typically leads to more significant renormalization group corrections. In the first case, all fields associated with the SM fermions are allowed to propagate in the bulk, whereas in the second, the matter fields are restricted to the brane. Neutrino masses evolve differently in the two models due to the sign of the (different) dominant contributions in the bulk and in the brane cases. For the brane case we find the approximate degenerate neutrino mass spectrum becomes more favourable at the ultraviolet cut-off, whilst in the bulk case, the neutrino splitting becomes even more severe as the unitarity bounds of the effective theory are rapidly reached. As the evaluation of RGEs may play a crucial role in searching for realistic mixing patterns we also studied the evolution of mixing angles and phases. Contrary to the large renormalization effect of $\theta_{12}$, the runnings of $\theta_{13}$ and $\theta_{23}$ were relatively mild. We found a non-zero value for $\theta_{13}$ during the evolution, which has no appreciable RGE running effects, even when power law evolution effects are considered. Therefore it is necessary to introduce new physics effects in order to achieve the tri-bimaximal pattern. Here, we also find the maximum CP violation case, $ \delta  = \frac{\pi }{2}$, could be achieved starting from a relatively small initial value. In this regard, radiative effects have a very significant impact on neutrino physics. A non-zero Jarlskog invariant, which measures the magnitude of leptonic CP violation and is expected to be measured in future long baseline neutrino oscillation experiments, could open the door for CP violation in the lepton sector.

\section*{Acknowledgements}

ASC would like to thank AD, AT and the IPNL for their hospitality during his stay in Lyon, where this work was completed. Also, our thanks to Sasha Davidson, for useful comments and discussions during the preparation of this paper.

\appendix


\section{Conventions for masses and mixing parameters}\label{app:0}

\par Within this appendix we would like to stress our conventions for the mixing angles and phases and briefly discuss different scenarios for neutrino masses. The mixing matrix which relates gauge and mass eigenstates is defined to diagonalise the neutrino mass matrix in the basis where the charged lepton mass matrix is diagonal. It is usually parameterised as follows~\cite{Maki:1962mu}:
\begin{equation}
U = \left( \begin{array}{ccc}
c_{12} c_{13} & s_{12} c_{13} & s_{13} e^{- i \delta} \\
-s_{12} c_{23} - c_{12} s_{23} s_{13} e^{i \delta} & c_{12} c_{23} - s_{12} s_{23} s_{13} e^{i \delta} & s_{23} c_{13} \\
s_{12} s_{23} - c_{12} c_{23} s_{13} e^{i \delta} & - c_{12} s_{23} - s_{12} c_{23} s_{13} e^{i \delta} & c_{23} c_{13}
\end{array} \right)
\left( \begin{array}{ccc}
e^{i \phi_1} && \\
& e^{i \phi_2} & \\
&& 1
\end{array} \right) \; , \nonumber
\end{equation}
with $c_{ij} = \cos \theta_{ij}$ and $s_{ij} = \sin \theta_{ij}$ ($ij = 12, 13, 23$).  We follow the conventions of Ref.~\cite{Antusch:2003kp} to extract mixing parameters from the PMNS matrix.

\par Experimental information on neutrino mixing parameters and masses is obtained mainly from oscillation experiments. In general $\Delta m^2_{\mathit{atm}}$ is assigned to a mass squared difference between $\nu_3$ and $\nu_2$, whereas $\Delta m^2_{\mathit{sol}}$ to a mass squared difference between $\nu_2$ and $\nu_1$. The current observational values are summarised in Table~\ref{tableexp}. Data indicates that $\Delta m^2_{\mathit{sol}} \ll \Delta m^2_{\mathit{atm}}$, but the masses themselves are not determined. In this work we have adopted the masses of the neutrinos at the $M_Z$ scale as $m_{1} = 0.1$ eV, $m_{2} = 0.100379$ eV, and $m_{3} = 0.11183$ eV, as the {\it normal} hierarchy (whilst any reference to an {\it inverted} hierarchy would refer to $m_{3} = 0.1$ eV, with $m_{3} < m_{1} < m_{2}$ and satisfying the above bounds). For the purpose of illustration, we choose values for the angles and phases as the $M_Z$ scale as: $\theta_{12} = 34^o$, $\theta_{13} = 8.83^o$, $\theta_{23} = 46^o$, $\delta =30^0$, $\phi_1 = 80^o$ and $\phi_2 = 70^o$.

\begin{center}
\begin{table}[h!]
\begin{tabular}{c|c}
Parameter & Value (90\% CL) \\ \hline
$\sin^2(2\theta_{12})$ & $0.861(^{+0.026}_{-0.022})$ \\
$\sin^2(2\theta_{23})$ & $>0.92$ \\
$\sin^2(2\theta_{13})$ & $0.092\pm0.017 $ \\
$\Delta m^2_{\mathit{sol}}$ & $(7.59\pm 0.21)\times 10^{-5}eV^2$ \\
$\Delta m^2_{\mathit{atm}}$ & $(2.43\pm 0.13)\times 10^{-3}$ $eV^2$ \\
\end{tabular}
\caption{\it Present limits on neutrino masses and mixing parameters used in the text. Data is taken from Ref.~\protect{\cite{An:2012eh}} for $\sin^2(2\theta_{13})$, and from Ref.~\protect{\cite{Nakamura:2010zzi}}.}
\label{tableexp}
\end{table}
\end{center}


\section{Yukawa couplings}\label{app:A}

\par In the following we write for completeness all the evolution equations, where we shall use a notation similar to the ones of Refs.~\cite{Antusch:2003kp,Deandrea:2006mh}. Note that  the beta functions contain terms quadratic in the cut-off, where this part dominates the evolution of the Yukawa couplings and of $k$. The top Yukawa coupling becomes non-perturbative before the gauge couplings thus limiting the range of validity of the effective theory. We shall first write down the result of the MSSM and then generalize it to the inclusion of the effects arising from the extra dimensional degrees of freedom.

\par As in our previous works, the initial values we shall adopt at the $M_Z$ scale are: for the gauge couplings we have $\alpha_1(M_Z) = 0.01696$, $\alpha_2(M_Z) = 0.03377$, and $\alpha_3(M_Z) = 0.1184$, and for the fermion masses $m_u(M_Z) = 1.27$ MeV, $m_c(M_Z) = 0.619$ GeV, $m_t (M_Z) = 171.7$ GeV, $m_d(M_Z) = 2.90$ MeV, $m_s(M_Z) = 55$ MeV, $m_b (M_Z) = 2.89$ GeV, $m_e(M_Z) = 0.48657$ MeV, $m_\mu(M_Z) = 102.718$ MeV, and $m_\tau(M_Z) = 1746.24$ MeV \cite{Xing:2007fb,Cornell:2011fw}.

\subsection{MSSM}

\par The evolution equations for the MSSM are a limiting case of the 5D models we shall consider in the following. In any case, when $0 < t < \ln (\frac{1}{{{M_Z}R}})$ (that the energy we consider for the evolution is ${M_Z} < \mu < 1/R$) the Yukawa evolution equations are dictated by the usual MSSM:
\begin{eqnarray}
16{\pi ^2}\frac{{d{Y_d}}}{{dt}} &=& {Y_d}(3Tr(Y_d^\dag {Y_d}) + Tr(Y_e^\dag {Y_e}) + 3Y_d^\dag {Y_d} + Y_u^\dag {Y_u}) - {Y_d} \left( {\frac{7}{{15}}g_1^2 + 3g_2^2 + \frac{{16}}{3}g_3^2} \right) \; ,\nonumber\\
16{\pi ^2}\frac{{d{Y_u}}}{{dt}} &=& {Y_u}(3Tr(Y_u^\dag {Y_u}) + 3Y_u^\dag {Y_u} + Y_d^\dag {Y_d}) - {Y_u}\left( {\frac{{13}}{{15}}g_1^2 + 3g_2^2 + \frac{{16}}{3}g_3^2} \right) \; , \\
16{\pi ^2}\frac{{d{Y_e}}}{{dt}} &=& {Y_e}(3Tr(Y_d^\dag {Y_d}) + Tr(Y_e^\dag {Y_e}) + 3Y_e^\dag {Y_e}) - {Y_e}\left( {\frac{9}{5}g_1^2 + 3g_2^2} \right)\; . \nonumber
\end{eqnarray}
These equations are modified when we enter the energy regime where the effects of the extra dimensions sets in. The modifications depend on the particles non-decoupled at a certain energy and on the structure of the model. We shall consider two cases, one in which all particles can propagate in the extra dimensions (bulk scenario) and the other in which SM particles are constrained to the brane (brane scenario).

\subsection{The Bulk Scenario}

\par The RGEs for the Yukawa couplings in the 5D MSSM, for all three generations propagating in the bulk, can be expressed as:
\begin{eqnarray}
16{\pi ^2}\frac{{d{Y_d}}}{{dt}} &=& {Y_d}(3Tr(Y_d^\dag {Y_d}) + Tr(Y_e^\dag {Y_e}) + 3Y_d^\dag {Y_d} + Y_u^\dag {Y_u})\pi S{(t)^2} - {Y_d}\left( \frac{7}{15} g_1^2 + 3 g_2^2 + \frac{16}{3} g_3^2 \right) S(t) \; ,  \nonumber\\
16{\pi ^2}\frac{{d{Y_u}}}{{dt}} &=& {Y_u}(3Tr(Y_u^\dag {Y_u}) + 3Y_u^\dag {Y_u} + Y_d^\dag {Y_d})\pi S{(t)^2} - {Y_u}\left( {\frac{{13}} {{15}}g_1^2 + 3g_2^2 + \frac{{16}}{3}g_3^2} \right)S(t) \; ,  \\
16{\pi ^2}\frac{{d{Y_e}}}{{dt}} &=& {Y_e}(3Tr(Y_d^\dag {Y_d}) + Tr(Y_e^\dag {Y_e}) + 3Y_e^\dag {Y_e})\pi S{(t)^2} - {Y_e}\left( {\frac{9} {5}g_1^2 + 3g_2^2} \right)S(t) \;\nonumber
\end{eqnarray}
That is, when the energy scale $\mu > 1/R$ or when the energy scale parameter $t > \ln (\frac{1}{{{M_Z}R}})$ (where we have set ${M_Z}$ as the renormalization point, and use $S(t) = {e^t}{M_Z}R$). We can convert these equations to the following form:
\begin{eqnarray}
16{\pi ^2}\frac{{d{Y_d}}}{{dt}} &=& {Y_d}\left\{ {{T_d}\pi {S^2} - {G_d} + (3Y_d^\dag {Y_d} + Y_u^\dag {Y_u})\pi {S^2}} \right\} \; , \nonumber\\
16{\pi ^2}\frac{{d{Y_u}}}{{dt}} &=& {Y_u}\left\{ {{T_u}\pi {S^2} - {G_u} + (3Y_u^\dag {Y_u} + Y_d^\dag {Y_d})\pi {S^2}} \right\} \; , \\
16{\pi ^2}\frac{{d{Y_e}}}{{dt}} &=& {Y_e}\left\{ {{T_e}\pi {S^2} - {G_e} + (3Y_e^\dag {Y_e})\pi {S^2}} \right\} \; , \nonumber
\end{eqnarray}
where
\begin{eqnarray}
{T_d}&=&3\; Tr(Y_d^\dag {Y_d}) + Tr(Y_e^\dag {Y_e})\; , \nonumber\\
{G_d}&=&\left( {\frac{7}{{15}}g_1^2 + 3g_2^2 + \frac{{16}}{3}g_3^2} \right)S(t)\; , \nonumber\\
{T_u}&=&3\; Tr(Y_u^\dag {Y_u})\; ,\nonumber\\
{G_u}&=&\left( {\frac{{13}}{{15}}g_1^2 + 3g_2^2 + \frac{{16}}{3}g_3^2} \right)S(t)\; , \\
{T_e}&=&3\; Tr(Y_d^\dag {Y_d}) + Tr(Y_e^\dag {Y_e})\; ,\nonumber\\
{G_e}&=&\left( {\frac{9}{5}g_1^2 + 3g_2^2} \right)S(t)\; .\nonumber
\end{eqnarray}

\subsection{The Brane Scenario}

\par The results for the beta function in the case where all matter superfields are constrained to live on the 4D brane, the quadratic evolution due to the sum over two Kaluza-Klein towers will be milder. The evolution equations for the Yukawa couplings are given by:
\begin{eqnarray}
16{\pi ^2}\frac{{d{Y_d}}}{{dt}} &=& {Y_d}(3Tr(Y_d^\dag {Y_d}) + Tr(Y_e^\dag {Y_e}) + (6Y_d^\dag {Y_d} + 2Y_u^\dag {Y_u}) S{(t)}) - {Y_d} \left( \frac{19}{30} g_1^2 + \frac{9}{2} g_2^2 + \frac{32}{3} g_3^2 \right) S(t) \; , \nonumber\\
16{\pi ^2}\frac{{d{Y_u}}}{{dt}} &=& {Y_u}(3Tr(Y_u^\dag {Y_u}) + (6Y_u^\dag {Y_u} + 2Y_d^\dag {Y_d}) S{(t)}) - {Y_u}\left( {\frac{{43}}{{30}}g_1^2 + \frac{{9}}{{2}}g_2^2 + \frac{{32}}{3}g_3^2} \right)S(t) \; ,\\
16{\pi ^2}\frac{{d{Y_e}}}{{dt}} &=& {Y_e}(3Tr(Y_d^\dag {Y_d}) + Tr(Y_e^\dag {Y_e}) + 6Y_e^\dag {Y_e}S{(t)}) - {Y_e}\left( {\frac{33}{10}g_1^2 + \frac{9}{2}g_2^2} \right)S(t) \; . \nonumber
\end{eqnarray}
Which we can convert to the following:
\begin{eqnarray}
16{\pi ^2}\frac{{d{Y_d}}}{{dt}} &=& {Y_d}\left\{ {{T_d} - {G_{d}} + (6Y_d^\dag {Y_d} + 2Y_u^\dag {Y_u})S} \right\}\; , \nonumber\\
16{\pi ^2}\frac{{d{Y_u}}}{{dt}} &=& {Y_u}\left\{ {{T_u} - {G_{u}} + (6Y_u^\dag {Y_u} + 2Y_d^\dag {Y_d})S} \right\}\; , \\
16{\pi ^2}\frac{{d{Y_e}}}{{dt}} &=& {Y_e}\left\{ {{T_e} - {G_{e}} + (6Y_e^\dag {Y_e})S} \right\}\; , \nonumber
\end{eqnarray}
where
\begin{eqnarray}
{T_d}&=&3\; Tr(Y_d^\dag {Y_d}) + Tr(Y_e^\dag {Y_e})\; , \nonumber\\
{G_{d}}&=&\left( {\frac{19}{{30}}g_1^2 + \frac{9}{{2}}g_2^2 + \frac{{32}}{3}g_3^2} \right)S(t)\; ,\nonumber\\
{T_u}&=&3\; TrY_u^\dag {Y_u}\; , \nonumber\\
{G_{u}}&=&\left( {\frac{{43}}{{30}}g_1^2 + \frac{9}{{2}}g_2^2 + \frac{{32}}{3}g_3^2} \right)S(t)\; , \\
{T_e}&=&3\; Tr(Y_d^\dag {Y_d}) + Tr(Y_e^\dag {Y_e})\; , \nonumber\\
{G_e}&=&\left( {\frac{33}{10}g_1^2 +\frac{9}{{2}}g_2^2} \right)S(t)\; .\nonumber
\end{eqnarray}


\section{Mass and mixing angles evolution}\label{app:B}

\par As such, for the four dimensional MSSM, where $C=1$ and $\alpha=6Tr(Y_u^\dag {Y_u}) - \frac{6}{5}g_1^2 - 6 g_2^2 = 6(y_t^2 + y_c^2 +y_u^2)- \frac{6}{5}g_1^2 - 6 g_2^2$, we have:
\begin{eqnarray}
16{\pi ^2}\frac{{d{m_1}}}{{dt}} &=& {m_1}\left\{ \alpha + C y_{\tau}^2 \left( 2\sin^2(\theta_{12}) \sin^2(\theta_{23})-\sin(\theta_{13})\sin(2\theta_{12})\sin(2\theta_{23})\cos(\delta) \right. \right. \nonumber \\
&& \hspace{2cm} \left. \left. +2\sin^2(\theta_{13})\cos^2(\theta_{12})\cos^2(\theta_{23}) \right) \right\}\; , \nonumber\\
16{\pi ^2}\frac{{d{m_2}}}{{dt}} &=& {m_2}\left\{ \alpha + C y_{\tau}^2 \left( 2\cos^2(\theta_{12}) \sin^2(\theta_{23})+\sin(\theta_{13})\sin(2\theta_{12})\sin(2\theta_{23})\cos(\delta)\right. \right. \nonumber \\
&& \hspace{2cm} \left. \left. +2\sin^2(\theta_{13})\sin^2(\theta_{12})\cos^2(\theta_{23}) \right) \right\}\; , \nonumber\\
16{\pi ^2}\frac{{d{m_3}}}{{dt}} &=& {m_3}\left\{ \alpha + C y_{\tau}^2 \left( 2\cos^2(\theta_{13}) \cos^2(\theta_{23}) \right) \right\}\; ,
\end{eqnarray}
and for the mixing angles:
\begin{eqnarray}
16{\pi ^2}\frac{{d{\theta_{12}}}}{{dt}} &=& \frac{1}{2}C y_{\tau}^2 \left( \frac{m_1 + m_2}{m_1 - m_2}  \cos(\frac{\phi_1 - \phi_2}{2}) \Big( \frac{1}{2} \cos(\frac{\phi_1 - \phi_2}{2}) \cos^2(\theta_{13}) \sin(2\theta_{12})\right.\nonumber \\
&&+\left.\{ \frac{1}{4} \cos(\frac{\phi_1 - \phi_2}{2})(-3+ \cos(2\theta_{13})) \cos(2\theta_{23}) \sin(2\theta_{12})-[ \cos\delta  \cos(\frac{\phi_1 - \phi_2}{2}) \cos(2\theta_{12})\right.\nonumber \\
&&+ \left. \sin\delta  \sin(\frac{\phi_1 - \phi_2}{2})] \sin(\theta_{13}) \sin(2\theta_{23}) \} \Big)\right. \nonumber \\
&&+ \left.\frac{m_1 + m_3}{m_1 - m_3} \cos(\delta-\frac{\phi_1}{2}) \sin(\theta_{12}) \sin(\theta_{13}) \Big( - \cos(\frac{\phi_1} {2}) \sin(\theta_{12}) \sin(2\theta_{23}) \right. \nonumber \\
&&+ \left.  2 \cos(\delta-\frac{\phi_1}{2}) \cos(\theta_{12}) \sin(\theta_{13}) \cos^2(\theta_{23}) \Big)\right. \nonumber \\
&&+\left. \frac{m_2 + m_3}{m_2 - m_3} \cos(\delta-\frac{\phi_2}{2}) \cos(\theta_{12}) \tan(\theta_{13}) \Big(  \cos(\frac{\phi_2} {2}) \cos(\theta_{12}) \cos(\theta_{13}) \sin(2\theta_{23})\right. \nonumber \\
&&+ \left.  \cos(\delta-\frac{\phi_2}{2}) \sin(\theta_{12}) \sin(2\theta_{13}) \cos^2(\theta_{23}) \Big) \right)\; , \\
16{\pi ^2}\frac{{d{\theta_{13}}}}{{dt}} &=& \frac{1}{2}C y_{\tau}^2 \left( \frac{m_1 + m_3}{m_1 - m_3} \cos(\delta-\frac{\phi_1} {2}) \cos(\theta_{12}) \cos(\theta_{13}) \Big( - \cos(\frac{\phi_1}{2}) \sin(\theta_{12}) \sin(2\theta_{23})\right.\nonumber \\
&&+ \left.  2\cos(\delta-\frac{\phi_1}{2}) \cos(\theta_{12}) \sin(\theta_{13}) \cos^2(\theta_{23}) \Big) \right.\nonumber \\
&&+ \left. \frac{m_2 + m_3}{m_2 - m_3} \cos(\delta-\frac{\phi_2}{2}) \sin(\theta_{12}) \Big(  \cos(\frac{\phi_2} {2}) \cos(\theta_{12}) \cos(\theta_{13}) \sin(2\theta_{23})\right.\nonumber \\
&&+ \left.  \cos(\delta-\frac{\phi_2}{2}) \sin(\theta_{12}) \sin(2\theta_{13}) \cos^2(\theta_{23}) \Big) \right)\; , \\
16{\pi ^2}\frac{{d{\theta_{23}}}}{{dt}} &=& \frac{1}{2}C y_{\tau}^2 \left( \frac{m_1 + m_3}{m_1 - m_3} \cos(\frac{\phi_1}{2}) \sin(\theta_{12}) \Big(  \cos(\frac{\phi_1}{2}) \sin(\theta_{12}) \sin(2\theta_{23}) \right. \nonumber \\
&& \left. -2 \cos(\delta-\frac{\phi_1}{2}) \cos(\theta_{12}) \sin(\theta_{13}) \cos^2(\theta_{23}) \Big) \right.\nonumber \\
&&+\left. \frac{m_2 + m_3}{m_2 - m_3} \cos(\frac{\phi_2}{2}) \cos(\theta_{12}) \sec(\theta_{13}) \Big(  \cos(\frac{\phi_2} {2}) \cos(\theta_{12}) \cos(\theta_{13}) \sin(2\theta_{23})\right.\nonumber \\
&&+ \left.  \cos(\delta-\frac{\phi_2}{2}) \sin(\theta_{12}) \sin(2\theta_{13}) \cos^2(\theta_{23}) \Big) \right)\; ,
\end{eqnarray}

\par As we have already mentioned, the transition to the bulk case will be done by making the replacement of $C=C(\mu)=\pi \mu^2 R^2=\pi S(t)^2$ and $\alpha=6\pi S(t)^2 Tr(Y_u^\dag {Y_u}) - (\frac{6}{5}g_1^2 +6 g_2^2)S(t)$. Similarly, we will also have the same equations in the brane case, and with $C=C(\mu)=2\mu R=2S(t)$ and $\alpha=6Tr(Y_u^\dag {Y_u}) - (\frac{9}{5}g_1^2 + 9 g_2^2)S(t)$.


\section{Dirac and Majorana phases evolution}\label{app:C}

\par As already discussed, in the MSSM $C=1$, in the 5D brane case $C=2 \mu R=2 S(t)$, and in the bulk case $C=C(\mu)=\pi \mu^2 R^2=\pi S(t)^2$. In which case, the equations for the evolution of the phases can be expressed as:
\begin{eqnarray}
16\pi^2 \frac{d\phi_1}{dt} &=& \frac{1}{2}C y_{\tau}^2 \left( \frac{1}{2}\frac{m_1 + m_2}{m_1 - m_2}  \sin(\frac{\phi_1 - \phi_2}{2})
\cot(\theta_{12})\Big( -2 \cos(\frac{\phi_1 - \phi_2}{2}) \cos^2(\theta_{13}) \sin(2\theta_{12})\right.\nonumber \\
&-&\left.  \Big\{  \cos(\frac{\phi_1 - \phi_2}{2})(-3+ \cos(2\theta_{13})) \cos(2\theta_{23}) \sin(2\theta_{12})\right.\nonumber \\
&-& \left. 4 \left[ \cos\delta  \cos(\frac{\phi_1 - \phi_2}{2}) \cos(2\theta_{12})+ \sin\delta  \sin(\frac{\phi_1 - \phi_2}{2}) \right]
\sin(\theta_{13}) \sin(2\theta_{23}) \Big\} \Big) \right.\nonumber\\
&+& \left. \frac{m_1 + m_3}{m_1 - m_3}2 \cos(\theta_{13}) \Big( - \cot(\theta_{23}) \sin(\frac{\phi_1}{2}) \sin(\theta_{12}) \sec(\theta_{13})
+2 \cos(\theta_{12}) \sin(\delta-\frac{\phi_1}{2}) \tan(\theta_{13})\right.\nonumber\\
&+&  \left. \sec(\theta_{13}) \tan(\theta_{23}) \sin(\frac{\phi_1}{2}) \sin(\theta_{12}) \Big) \Big( -2 \cos(\delta-\frac{\phi_1}{2})
\cos(\theta_{12}) \sin(\theta_{13}) \cos^2(\theta_{23})+ \cos(\frac{\phi_1}{2}) \sin(\theta_{12}) \sin(2\theta_{23}) \Big) \right. \nonumber \\
&+&  \left.\frac{m_2 + m_3}{m_2 - m_3} \Big(  \sin(\delta-\frac{\phi_2}{2}) \sin(\theta_{12}) \tan(\theta_{13})+ \cos(\theta_{12})
\Big\{  \cot(\theta_{23}) \sec(\theta_{13}) \sin(\frac{\phi_2}{2})\right. \nonumber \\
&-&\left. \cos(\frac{\phi_2}{2}) \cot(\theta_{12}) \sin\delta  \tan(\theta_{13})+\sin(\frac{\phi_2}{2}) \cos(\theta_{12}) \cos\delta  \tan(\theta_{13}) - \sec(\theta_{13}) \tan(\theta_{23})
\sin(\frac{\phi_2}{2}) \Big\} \Big) \right. \nonumber \\
&\times&\left. \Big( -2 \cos(\frac{\phi_2}{2}) \cos(\theta_{12}) \cos(\theta_{13}) \sin(2\theta_{23})
- 2 \cos(\delta-\frac{\phi_2}{2}) \sin(\theta_{12}) \sin(2\theta_{13}) \cos^2(\theta_{23}) \Big) \right)\; ,
\end{eqnarray}
\begin{eqnarray}
16\pi^2 \frac{d\phi_2}{dt} &=& \frac{1}{2}C y_{\tau}^2 \left( \frac{1}{2}\frac{m_1 + m_2}{m_1 - m_2}  \sin(\frac{\phi_1 - \phi_2}{2})
\tan(\theta_{12})\Big( -2 \cos(\frac{\phi_1 - \phi_2}{2}) \cos^2(\theta_{13}) \sin(2\theta_{12})\right. \nonumber \\
&-&\left. \{  \cos(\frac{\phi_1 - \phi_2}{2})(-3+ \cos(2\theta_{13})) \cos(2\theta_{23}) \sin(2\theta_{12})-4[ \cos\delta
\cos(\frac{\phi_1 - \phi_2}{2}) \cos(2\theta_{12})\right. \nonumber \\
&+& \left. \sin\delta  \sin(\frac{\phi_1 - \phi_2}{2})] \sin(\theta_{13}) \sin(2\theta_{23}) \} \Big) \right. \nonumber \\
&-& \left.\frac{m_1 + m_3}{m_1 - m_3}2 \cos(\theta_{13}) \Big( \cot(\theta_{23}) \sin(\frac{\phi_1}{2}) \sin(\theta_{12})
\sec(\theta_{13})- \cos(\theta_{12}) \sin(\delta-\frac{\phi_1}{2}) \tan(\theta_{13})\right. \nonumber \\
&+&\left. \sin(\theta_{12}) \{  \sin\delta  \cos(\frac{\phi_1}{2}) \tan(\theta_{12})  \tan(\theta_{13})- \sin(\frac{\phi_1}{2})  \cos\delta  \tan(\theta_{12})  \tan(\theta_{13}) - \sin(\frac{\phi_1}{2}) \sec(\theta_{13}) \tan(\theta_{23}) \} \Big) \right. \nonumber \\
&\times&\left. \Big( -2 \cos(\delta-\frac{\phi_1}{2}) \cos(\theta_{12}) \sin(\theta_{13}) \cos^2(\theta_{23})+ \cos(\frac{\phi_1}{2}) \sin(\theta_{12}) \sin(2\theta_{23}) \Big) \right.\nonumber \\
&+& \left. \frac{m_2 + m_3}{m_2 - m_3} \Big(  \sin(\frac{\phi_2}{2}) \cos(\theta_{12}) \cos(2\theta_{23})\csc(\theta_{23}) \sec(\theta_{13}) \sec(\theta_{23})+2 \sin(\delta-\frac{\phi_2}{2}) \sin(\theta_{12}) \tan(\theta_{13}) \Big)  \right.\nonumber \\
&\times&\left. \Big( -2 \cos(\frac{\phi_2}{2}) \cos(\theta_{12}) \cos(\theta_{13}) \sin(2\theta_{23})-2 \cos(\delta-\frac{\phi_2}{2}) \sin(\theta_{12}) \sin(2\theta_{13}) \cos^2(\theta_{23}) \Big) \right)\; ,
\end{eqnarray}
\begin{eqnarray}
16\pi^2 \frac{d\delta}{dt} &=& \frac{1}{4}C y_{\tau}^2 \left( \frac{m_1 + m_2}{m_1 - m_2} \sin(\frac{\phi_1 - \phi_2}{2})\csc(2\theta_{12})\Big( -2\cos(\frac{\phi_1 - \phi_2}{2})\cos^2(\theta_{13})\sin(2\theta_{12})\right.\nonumber\\
&-&\left. \Big\{ \cos(\frac{\phi_1 - \phi_2}{2})(-3+\cos(2\theta_{13}))\cos(2\theta_{23})\sin(2\theta_{12})-4 [ \cos\delta \cos(\frac{\phi_1 - \phi_2}{2})\cos(2\theta_{12})\right.\nonumber\\
&+& \left. \sin\delta \sin(\frac{\phi_1 - \phi_2}{2}) ] \sin(\theta_{13})\sin(2\theta_{23}) \Big\} \Big) \right.\nonumber\\
&-&\left. \frac{m_1 + m_3}{8(m_1 - m_3)} \Big( \cos(\frac{\phi_1}{2}) \Big\{ 2-6\cos(2\theta_{12})
+  \cos(2\theta_{12}-2\theta_{13})-6\cos(2\theta_{13})\right.\nonumber\\
&+& \left. \cos(2\theta_{12}+2\theta_{13}) \Big\} \csc(\theta_{13})\sec(\theta_{12})\sin\delta+4\cos(\theta_{12})(-3+\cos(2\theta_{13}))
\csc(\theta_{13})\sin(\delta-\frac{\phi_1}{2})\right.\nonumber\\
&+& \left. \sin(\frac{\phi_1}{2}) \Big\{ -\cos(\delta) [ 2-6\cos(2\theta_{12})+\cos(2\theta_{12}-2\theta_{13})-6\cos(2\theta_{13})
+\cos(2\theta_{12}+2\theta_{13}) ] \csc(\theta_{13})\sec(\theta_{12})\right.\nonumber\\
&+& \left.16\sin(\theta_{12})\cot(\theta_{23})-16\sin(\theta_{12})\tan(\theta_{23}) \Big\} \Big) \Big( -2\cos(\delta-\frac{\phi_1}{2})\cos(\theta_{12})\sin(\theta_{13})\cos^2(\theta_{23})\right.\nonumber\\
&+& \left. \cos(\frac{\phi_1}{2})\sin(\theta_{12})\sin(2\theta_{23}) \Big) \right.\nonumber \\
&-& \left.\frac{m_2 + m_3}{4(m_2 - m_3)} \Big( [ \frac{-5}{2}+\frac{3}{2}\cos(2\theta_{12})-\frac{1}{4}\cos(2\theta_{12}-2\theta_{13})-\frac{1}{2}\cos(2\theta_{13})-\frac{1}{4}\cos(2\theta_{12}+2\theta_{13}) ] \right.\nonumber\\
&\times&\left. \csc(\theta_{12}) \csc(\theta_{13})\sec(\theta_{13})\sin(\delta-\frac{\phi_2}{2})+ \cos(\theta_{12})\sec(\theta_{13}) [ -\cos(\frac{\phi_2}{2})(-3+\cos(2\theta_{13}))\cot(\theta_{12}) \csc(\theta_{13})\sin\delta \right.\nonumber\\
&-&\left. 4\sin(\frac{\phi_2}{2})\cot(\theta_{23})+\sin(\frac{\phi_2}{2})\cos\delta (-3+\cos(2\theta_{13}))\cot(\theta_{12})\csc(\theta_{13})+4\sin(\frac{\phi_2}{2})\tan(\theta_{23}) ] \Big)  \right.\nonumber\\
&\times&\left. \Big( -2\cos(\frac{\phi_2}{2})\cos(\theta_{12})\cos(\theta_{13})\sin(2\theta_{23})-2\cos(\delta-\frac{\phi_2}{2})\sin(\theta_{12})\sin(2\theta_{13})\cos^2(\theta_{23}) \Big) \right)\; .
\end{eqnarray}



\begin{thebibliography}{99}

\bibitem{Cornell:2011fw}
  A.~S.~Cornell, A.~Deandrea, L.~-X.~Liu and A.~Tarhini,
  Phys.\ Rev.\ D {\bf 85} (2012) 056001
  [arXiv:1110.1942 [hep-ph]].

\bibitem{Mohapatra:2005wg}
  R.~N.~Mohapatra, S.~Antusch, K.~S.~Babu, G.~Barenboim, M.~-C.~Chen, A.~de Gouvea, P.~de Holanda and B.~Dutta {\it et al.},
  Rept.\ Prog.\ Phys.\  {\bf 70} (2007) 1757
  [hep-ph/0510213].

\bibitem{Raidal:2008jk}
  M.~Raidal, A.~van der Schaaf, I.~Bigi, M.~L.~Mangano, Y.~K.~Semertzidis, S.~Abel, S.~Albino and S.~Antusch {\it et al.},
  Eur.\ Phys.\ J.\ C {\bf 57} (2008) 13
  [arXiv:0801.1826 [hep-ph]].

\bibitem{An:2012eh}
  F.~P.~An {\it et al.}  [DAYA-BAY Collaboration],
  Phys.\ Rev.\ Lett.\  {\bf 108} (2012) 171803
  [arXiv:1203.1669 [hep-ex]].

\bibitem{Ahn:2012nd}
  J.~K.~Ahn {\it et al.}  [RENO Collaboration],
Phys.\ Rev.\ Lett.\  {\bf 108}, 191802 (2012)
[arXiv:1204.0626 [hep-ex]].

\bibitem{Babu:1987im}
  K.~S.~Babu,
Z.\ Phys.\ C {\bf 35}, 69 (1987).

\bibitem{Liu:2009vh}
  L.~-X.~Liu,
Int.\ J.\ Mod.\ Phys.\ A {\bf 25}, 4975 (2010)
[arXiv:0910.1326 [hep-ph]].

\bibitem{Cornell:2010sz}
  A.~S.~Cornell and L.~-X.~Liu,
Phys.\ Rev.\ D {\bf 83}, 033005 (2011)
[arXiv:1010.5522 [hep-ph]].

\bibitem{Liu:2011gr}
  L.~-X.~Liu and A.~S.~Cornell,
PoS KRUGER {\bf 2010}, 045 (2010)
[arXiv:1103.1527 [hep-ph]].

\bibitem{Chankowski:1993tx}
  P.~H.~Chankowski and Z.~Pluciennik,
  Phys.\ Lett.\ B {\bf 316} (1993) 312
  [hep-ph/9306333].

\bibitem{Antusch:2001ck}
  S.~Antusch, M.~Drees, J.~Kersten, M.~Lindner and M.~Ratz,
  Phys.\ Lett.\ B {\bf 519} (2001) 238
  [hep-ph/0108005].

\bibitem{Chankowski:1999xc}
  P.~H.~Chankowski, W.~Krolikowski and S.~Pokorski,
  Phys.\ Lett.\ B {\bf 473} (2000) 109
  [hep-ph/9910231].

\bibitem{Casas:2003kh}
  J.~A.~Casas, J.~R.~Espinosa and I.~Navarro,
  JHEP {\bf 0309} (2003) 048
  [arXiv:hep-ph/0306243].

\bibitem{Blennow:2011mp}
  M.~Blennow, H.~Melbeus, T.~Ohlsson and H.~Zhang,
  JHEP {\bf 1104} (2011) 052
  [arXiv:1101.2585 [hep-ph]].

\bibitem{Antusch:2003kp}
  S.~Antusch, J.~Kersten, M.~Lindner and M.~Ratz,
  Nucl.\ Phys.\ B {\bf 674} (2003) 401
  [hep-ph/0305273].

\bibitem{Deandrea:2006mh}
  A.~Deandrea, J.~Welzel, P.~Hosteins and M.~Oertel,
  Phys.\ Rev.\ D {\bf 75} (2007) 113005
  [hep-ph/0611172].

\bibitem{Weinberg:1979sa}
  S.~Weinberg,
  Phys.\ Rev.\ Lett.\  {\bf 43} (1979) 1566.

\bibitem{Mirabelli:1997aj}
  E.~A.~Mirabelli and M.~E.~Peskin,
  Phys.\ Rev.\ D {\bf 58} (1998) 065002
  [hep-th/9712214].

\bibitem{ArkaniHamed:2001tb}
  N.~Arkani-Hamed, T.~Gregoire and J.~G.~Wacker,
  JHEP {\bf 0203} (2002) 055
  [hep-th/0101233].

\bibitem{Hebecker:2001ke}
  A.~Hebecker,
  Nucl.\ Phys.\ B {\bf 632} (2002) 101
  [hep-ph/0112230].

\bibitem{Flacke:2003ac}
  T.~Flacke,
  DESY-THESIS-2003-047.

 \bibitem{Harrison:2002er}
  P.~F.~Harrison, D.~H.~Perkins and W.~G.~Scott,
Phys.\ Lett.\ B {\bf 530}, 167 (2002)
[hep-ph/0202074].

\bibitem{Buchmuller:2003gz}
  W.~Buchmuller, P.~Di Bari and M.~Plumacher,
  Nucl.\ Phys.\ B {\bf 665} (2003) 445
  [hep-ph/0302092].

\bibitem{Gu:2010ye}
  P.~-H.~Gu,
  Phys.\ Rev.\ D {\bf 81} (2010) 073002
  [arXiv:1001.1340 [hep-ph]].

\bibitem{Luo:2012ce}
  S.~Luo and Z.~-z.~Xing,
arXiv:1203.3118 [hep-ph].

\bibitem{Maki:1962mu}
  Z.~Maki, M.~Nakagawa and S.~Sakata,
  Prog.\ Theor.\ Phys.\  {\bf 28} (1962) 870.

\bibitem{Nakamura:2010zzi}
  K.~Nakamura {\it et al.}  [Particle Data Group Collaboration],
  J.\ Phys.\ G G {\bf 37} (2010) 075021 and 2011 partial update for the 2012 edition.

\bibitem{Xing:2007fb}
  Z.~-z.~Xing, H.~Zhang and S.~Zhou,
  Phys.\ Rev.\ D {\bf 77}, 113016 (2008)
  [arXiv:0712.1419 [hep-ph]].

 \end{thebibliography}
\end{document}